\documentclass[aps,10pt,showpacs,preprint,longbibliography, superscriptaddress, english, prx]{revtex4-2}
\usepackage[T1]{fontenc}
\usepackage[ruled,linesnumbered]{algorithm2e}
\usepackage{dsfont}
\usepackage{amsmath}
\usepackage{amssymb}
\DeclareMathOperator{\sign}{sign}

\usepackage{bm}

\usepackage{xspace}
\newcommand*{\eg}{\emph{e.g.}\@\xspace}
\newcommand*{\ie}{\emph{i.e.}\@\xspace}

\usepackage{enumitem}
\usepackage{graphicx}
\graphicspath{{figs/}}
\usepackage{subcaption}
\usepackage[justification=raggedright]{caption}

\usepackage[colorlinks=true,urlcolor=blue,citecolor=blue,linkcolor=blue]{hyperref} 

\usepackage[version=4]{mhchem}
\usepackage{mathtools}
\usepackage{siunitx}

\makeatletter

\begin{document}

\title{Differentiable free energy surface:
a variational approach to directly observing rare events using generative deep-learning models}
\author{Shuo-Hui Li}
\email{shuohuili@ust.hk}
\affiliation{Department of Physics, The Hong Kong University of Science and Technology, Hong Kong, China}

\author{Chen Chen}
\affiliation{Department of Physics, The Hong Kong University of Science and Technology, Hong Kong, China}

\author{Yao-Wen Zhang}
\affiliation{Department of Physics, The Hong Kong University of Science and Technology, Hong Kong, China}

\author{Ding Pan}
\email{dingpan@ust.hk}
\affiliation{Department of Physics, The Hong Kong University of Science and Technology, Hong Kong, China}
\affiliation{Department of Chemistry, The Hong Kong University of Science and Technology, Hong Kong, China}
\affiliation{IAS Center for AI for Scientific Discoveries, Hong Kong University of Science and Technology, Hong Kong, China}

\begin{abstract}
Rare events are central to the evolution of complex many-body systems, characterized as key transitional configurations on the free energy surface (FES).
Conventional methods require adequate sampling of rare event transitions to obtain the FES, which is computationally very demanding.
Here we introduce the variational free energy surface (VaFES), a dataset-free framework that directly models FESs using tractable-density generative models.
Rare events can then be immediately identified from the FES with their configurations generated directly via one-shot sampling of generative models.
By extending a coarse-grained collective variable (CV) into its reversible equivalent, VaFES constructs a latent space of intermediate representation in which the CVs explicitly occupy a subset of dimensions.
This latent-space construction preserves the physical interpretability and transparent controllability of the CVs by design, while accommodating arbitrary CV formulations.
The reversibility makes the system energy exactly accessible, enabling variational optimization of the FES without pre-generated simulation data.
A single optimization yields a continuous, differentiable FES together with one-shot generation of rare-event configurations.
Our method can reproduce the exact analytical solution for the bistable dimer potential and identify a chignolin native folded state in close alignment with the experimental NMR structure.
Our approach thus establishes a scalable, systematic framework for advancing the study of complex statistical systems.
\end{abstract}

\maketitle

\section{Introduction}

Rare events often govern the key processes of complex statistical systems, ranging from chemical reactivity and biomolecular function to seismic activity.
In atomistic simulations, these intrinsically slow processes require trajectories to traverse vast high-dimensional configuration spaces and cross high kinetic barriers between metastable basins, making sampling very costly: molecular dynamics (MD) may require exhaustive simulations reaching milliseconds to seconds or longer, while Markov chain Monte Carlo (MCMC) can demand prohibitively many accept--reject steps.
In practice, rare-event sampling is therefore typically performed in a reduced subspace of coarse-grained collective variables (CVs), which paves the foundation for influential computational methods such as umbrella sampling~\cite{umbrellaSampling}, thermodynamic integration~\cite{TI1, TI2}, the Wang--Landau algorithm~\cite{WangLandau}, and metadynamics~\cite{metaD}.
Despite operating in this coarse-grained subspace, these methods still face convergence difficulties due to high barriers, low event probabilities, and technical issues such as poor CV choices or sampling parameter selections~\cite{metaDreview1, metaDreview2}.
The concept of the free energy surface (FES) follows naturally once all rare-event transitions associated with the CVs are identified and sufficiently sampled.
Formally, through histogramming and reweighting, the FES value $F(\bm{s})$ at a given CV coordinate $\bm{s}$ is obtained by integrating over the corresponding configurations:
\begin{equation}
  F(\bm{s}) = -T\ln \int\ \exp\left(-\beta E(\bm{x})\right) \delta(\bm{s} - \mathcal{S}(\bm{x})) d\bm{x},
  \label{eq:FES1}
\end{equation}
where $\beta = 1/T$, $\bm{x}$ denotes a configuration with energy $E(\bm{x})$, and the coarse-graining map $\mathcal{S}$ converts $\bm{x}$ into the corresponding CV coordinate $\bm{s}$.
The FES emerges as an effective landscape governing the coarse-grained dynamics, therefore playing a central role in identifying metastable states, quantifying transition barriers, and elucidating reaction pathways.
Nevertheless, because even a single rare-event transition is difficult to sample and converge, accurate estimation of the full FES remains exceptionally challenging for sampling-based methods.

Variational methods offer a distinct alternative by targeting free energies directly~\cite{meanfield}.
Recent advances have employed tractable-density generative models~\cite{glow, realnvp, pixelRCNN, frankNoe} as highly expressive variational models capable of one-shot direct sampling from the target distribution~\cite{prl, wuDianPRL, prr, vatd}.
These approaches have demonstrated substantial success in estimating free energies for complex systems, including lattice models~\cite{vatd, prl}, classical atomistic systems~\cite{atomicNF1, atomicNF2, prx}, and quantum atomic systems~\cite{LWnewPRL}.
Remarkably, such neural networks exhibit superior scalability to high dimensionality, as routinely demonstrated in demanding machine-learning applications involving millions to billions of parameters~\cite{frankNoe, zhangLFDP, diffusionModel, chatGPT, chemDL, chemDL2, genAI4chem, genAI4chem2, genAI4chem3}.
In principle, the FES should admit a similar variational treatment~\cite{vfep, PRLvfe}, yet a key obstacle remains: existing deep-learning variational formulations require the exact energy function in the original full configuration space, which is no longer accessible after coarse-graining onto CVs.
A framework that starts from the energy function and directly yields the FES as a continuous function of the CVs therefore remains an open challenge.

Here we introduce the variational free energy surface (VaFES), a framework for directly modeling FESs with tractable-density generative models, in which the FES is first obtained as a characterizing map and then used to identify and generate rare-event configurations via one-shot sampling.
The key idea underlying the variational formulation is to replace the coarse-grained CV transformation with a bijection.
Exploiting the universality of bijective transformations---a concept that also underlies the universality of quantum computing~\cite{nielsen2010quantum}---we expand any given coarse-grained CV transformation into its reversible equivalent.
The resulting construction embeds the CV coordinates into an intermediate representation with the same dimensionality as the original full space.
Leveraging this reversibility, the system energy can be rewritten exactly, enabling variational optimization of the FES.
With no need for external datasets or time-consuming MCMC/MD simulations in the optimization loop,
the generative model is trained on self-generated one-shot configurations in a bootstrap manner.
After training, the variational objective yields the FES as a continuous, differentiable function of the CVs, in contrast to the histogrammed, discrete free energy profiles produced by conventional deep-learning approaches~\cite{frankNoe, conditionalBG, conditionedBGtps, enhancedDiffusion}.
This allows rare-event transitions to be identified using methods such as the nudged elastic band (NEB)~\cite{neb1, neb2} and the string method~\cite{stringMethod}.
Subsequently, direct sampling from the model yields one-shot configurations conditioned on the identified CV values.
A single optimization therefore provides both a continuous, differentiable FES and the corresponding rare-event configurations.

The intermediate representation in VaFES combines physical interpretability with transparent controllability.
By construction, the CVs explicitly occupy a subset of latent dimensions, therefore carrying clear physical or chemical meaning, unlike the opaque latent spaces produced by stochastic optimization~\cite{vae, gan, realnvp, diffusionModel, frankNoe}.
The design also remains flexible enough to incorporate machine-learned or statistically derived CV coordinates.
Furthermore, the universality of bijections removes structural restrictions on CV choice, allowing VaFES to work with arbitrary CV formulations beyond the limitations of conventional methods~\cite{umbrellaSampling, TI1, TI2}.
At the same time, it enables systematic symmetry removal and the imposition of exact geometric constraints, providing an alternative to architectural inductive biases typically imposed by equivariant neural networks~\cite{equivariant1, equivariant2, equivariantNet, winter2022unsupervised}.

We demonstrate VaFES across applications of increasing complexity, from an analytically solvable toy model to atomistic systems, concluding in the protein chignolin.
In the toy-model benchmark of Sec.~\ref{sec:dimmer}, VaFES reproduces the analytical solution exactly.
In Sec.~\ref{sec:h2n2} and Sec.~\ref{sec:aldp}, the intermediate representation integrates physical and data-driven CVs seamlessly, while local-frame constructions eliminate translational and rotational degrees of freedom.
Finally, in the chignolin application of Sec.~\ref{sec:chignolin}, we employ a more sophisticated intermediate representation that converts all Cartesian coordinates into physics-informed latent dimensions, allowing exact geometric constraints such as bond-length upper bounds, L-type chirality, and planar aromatic rings represented in two-dimensional local frames.
The resulting native folded state prediction achieves a \ce{C_{\alpha}} root-mean-square deviation (\ce{C_\alpha}-RMSD) of approximately $1\si{\angstrom}$ from the experimental NMR structure, highlighting the ability of VaFES to recover realistic rare-event configurations with high structural fidelity.

\section{Results and discussion}

\subsection{Variational free energy surface}\label{sec:vafes}
\begin{figure*}[!htbp]
\centering
\includegraphics[width=0.7\textwidth]{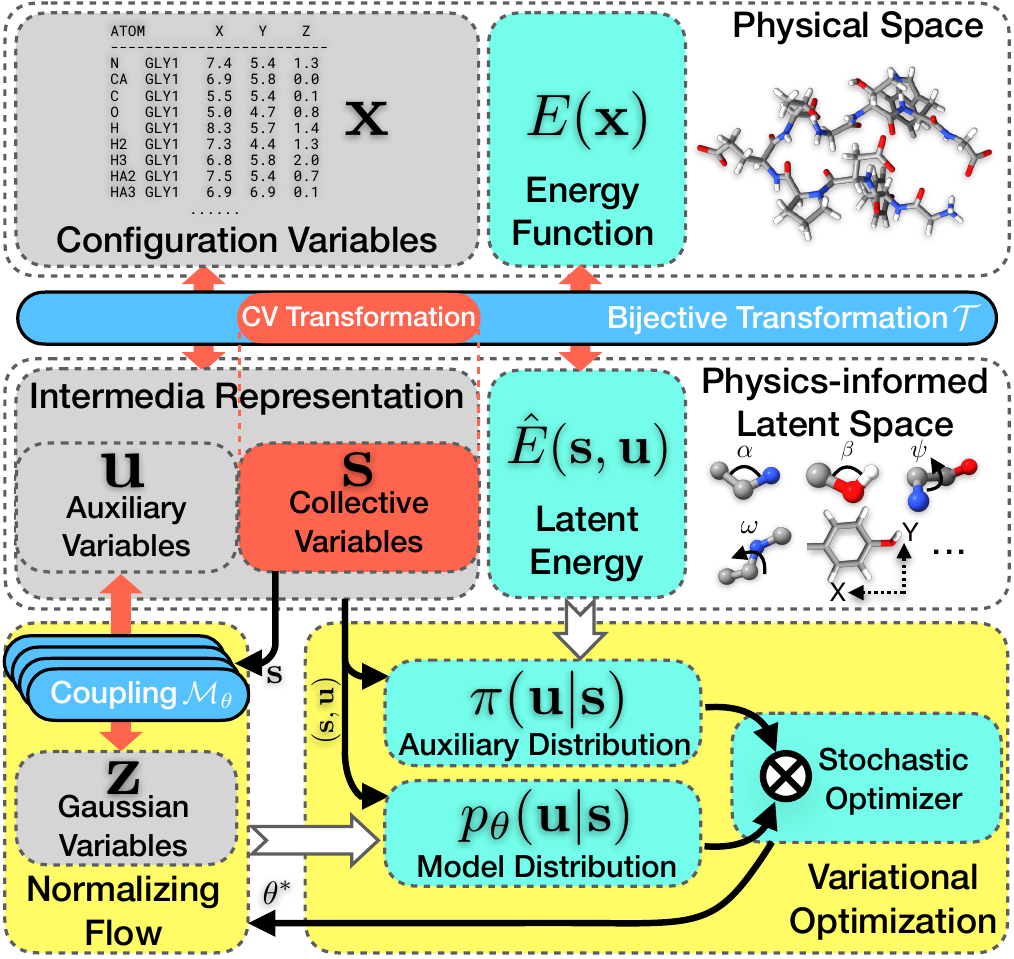}
\caption{
  Schematic illustration of the VaFES framework.
  The coarse-grained collective variable (CV) transformation is expanded into a bijective transformation $\mathcal{T}$, simultaneously establishing two reversible mappings.
  It maps configuration variables of physical space, $\bm{x}$, to the intermediate representation $(\bm{s}, \bm{u})$, where the CVs $\bm{s}$ explicitly occupy a subset of the dimensions and the remaining degrees of freedom are collected in the auxiliary variables $\bm{u}$.
  The same $\mathcal{T}$ transformation also maps the physical energy function into the latent-space energy of intermediate representations, thereby defining the conditional distribution $\pi(\bm{u} | \bm{s})$ for the auxiliary variables $\bm{u}$.
  A tractable-density generative model, illustrated here by a multi-coupling normalizing flow, parameterizes $p_{\theta}(\bm{u} | \bm{s})$ and enables one-shot direct sampling of auxiliary variables conditioned on the CVs.
  Minimizing the divergence between $p_{\theta}(\bm{u} | \bm{s})$ and $\pi(\bm{u} | \bm{s})$ yields a variational estimate of the free energy surface and, through the inverse map $\mathcal{T}^{-1}$, the corresponding physical configurations.
}
\label{fig:diagram}
\end{figure*}

In this section, we describe the detailed design of the VaFES framework, schematically illustrated in Fig.~\ref{fig:diagram}.
In VaFES, we first reformulate the coarse-grained CV transformation $\mathcal{S}$ as a reversible equivalent $\mathcal{T}$, exploiting the fact that any transformation can be extended to a corresponding bijection~\cite{reversibleComp2, reversibleCompBook}.
For a given input configuration $\bm{x}$, the bijective transformation $\mathcal{T}$ returns the original coarse-grained variables, the CVs $\bm{s}$, together with additional dimensions, denoted by the auxiliary variables $\bm{u}$, \ie,
$(\bm{s}, \bm{u}) = \mathcal{T}(\bm{x})$.
We refer to the pair $(\bm{s}, \bm{u})$ as the intermediate representation,
which preserves all information in the original configuration $\bm{x}$ and therefore admits an inverse transformation.
An immediate advantage of this representation is that, in the latent space spanned by $(\bm{s}, \bm{u})$, the FES in Eq.~(\ref{eq:FES1}) can be rewritten in a clear form
\begin{equation}
  F(\bm{s}) = -T\ln \int\ \exp\left(-\beta \hat{E}(\bm{s}, \bm{u})\right) d\bm{u},
  \label{eq:FES2}
\end{equation}
where the integration is performed only over the auxiliary variables $\bm{u}$.
The corresponding energy function of the intermediate representation, $\hat{E}(\bm{s}, \bm{u})$, follows directly from reversibility and the change-of-variables formula:
\begin{equation}
  \hat{E}(\bm{s}, \bm{u}) = E(\mathcal{T}^{-1}(\bm{s}, \bm{u})) - T \ln\left|\det\frac{\partial \mathcal{T}^{-1}}{\partial (\bm{s}, \bm{u})}\right|,
  \label{eq:CVE}
\end{equation}
where $\frac{\partial \mathcal{T}^{-1}}{\partial (\bm{s}, \bm{u})}$ is the Jacobian matrix of the inverse transformation $\mathcal{T}^{-1}$.

As a canonical ensemble, the energy function in Eq.~(\ref{eq:CVE}) admits a Boltzmann distribution over the intermediate representation $(\bm{s}, \bm{u})$.
Given a target CV value $\bm{s}$, Eq.~(\ref{eq:CVE}) can be regarded as an energy function of $\bm{u}$ alone, thereby defining the \emph{conditional} Boltzmann distribution of the auxiliary variables:
\begin{equation}
  \pi(\bm{u} | \bm{s}) = \frac{\exp\left(-\beta \hat{E}(\bm{s}, \bm{u})\right)}{\mathcal{Z}(\bm{s})},
  \label{eq:distribution}
\end{equation}
where $\mathcal{Z}(\bm{s}) = \int \exp(-\beta \hat{E}(\bm{s}, \bm{u})) d\bm{u}$, the negative logarithm of which is exactly Eq.~(\ref{eq:FES2}) without the temperature factor $T$.
To estimate $\mathcal{Z}(\bm{s})$, we introduce a variational model $p_{\theta}(\bm{u} | \bm{s})$, parameterized by $\theta$, and minimize the Kullback-Leibler divergence (KLD)~\cite{KLD},
\begin{equation}
  D_{KL}\Big(p_{\theta}(\bm{u} | \bm{s})|| \pi(\bm{u} | \bm{s}) \Big) =  \ln \mathcal{Z}(\bm{s}) + \underset{\bm{u}\sim p_{\bm{s}, \theta}}{\mathbb{E}} \left[\ln p_{\theta}(\bm{u} | \bm{s}) + \beta \hat{E}(\bm{s}, \bm{u})\right].
  \label{eq:kld}
\end{equation}
Because the KLD is always non-negative, \ie, $D_{KL} \ge 0$, the stochastic estimator
$T\underset{\bm{u}\sim p_{\bm{s}, \theta}}{\mathbb{E}} [\ln p_{\theta}(\bm{u} | \bm{s}) + \beta \hat{E}(\bm{s}, \bm{u})]$
provides an upper bound, denoted by $\bar{F}(\bm{s})$, for the exact FES value $F(\bm{s})$.
Minimizing the KLD therefore yields a variational estimate of the FES.

Next, rather than optimizing Eq.~(\ref{eq:kld}) separately at each CV point, we aim to access the FES over the entire CV space within a single optimization.
By treating the CVs as external parameters in the Boltzmann distribution of the auxiliary variables (Eq.~(\ref{eq:distribution})), analogous to the inverse temperature $\beta$, we can adopt the variational optimization proposed in~\cite{vatd} to estimate the FES as a continuous function of the CVs.
This leads to the loss function
\begin{equation}
   \underset{\theta}{\text{min}} \underset{\bm{s}\sim U, \bm{u}\sim p_{\bm{s}, {\theta}}}{\mathbb{E}} \left[T \ln p_{\theta}(\bm{u} | \bm{s}) + \hat{E}(\bm{s}, \bm{u}) \right] = \underset{\theta}{\text{min}} \underset{\bm{s}\sim U}{\mathbb{E}}\bar{F}(\bm{s}), 
  \label{eq:loss}
\end{equation}
where $U$ denotes a uniform distribution.

As in standard variational methods, this objective requires evaluation of the normalized model density $p_{\theta}(\bm{u} | \bm{s})$.
Efficient optimization of Eq.~(\ref{eq:loss}) further requires one-shot direct sampling from the variational model, making deep generative models a natural choice.
Accordingly, VaFES adopts a tractable-density generative model as its variational model.
Because any such model can both evaluate normalized densities and generate one-shot samples directly, the framework is not tied to a specific architecture.
For demonstration, in this section and in the applications below, we use a cubic-spline normalizing flow (cubic-spline flow)~\cite{cubic}.

Because the auxiliary variables $\bm{u}$ are sampled directly from $p_{\theta}(\bm{u} | \bm{s})$, this optimization requires no pre-generated dataset.
By contrast, conventional deep-learning generative approaches~\cite{DiG, BioEmu, alphaFold3, simpleFold} rely on large pre-generated datasets, making their performance sensitive to data quality and coverage; the resulting ensembles inevitably inherit biases present in the training data.
VaFES avoids these limitations by starting directly from the energy function and formulating the FES variationally as a continuous function of the CVs.
After optimization, for any specified CV value $\bm{s}$, Eq.~(\ref{eq:loss}) yields the FES estimate $\bar{F}(\bm{s})$, which is continuous and differentiable with respect to $\bm{s}$ and closely resembles an analytical solution.
This differs from conventional deep-learning generative approaches~\cite{frankNoe, conditionalBG, conditionedBGtps, enhancedDiffusion}, which typically recover free energy profiles only indirectly through reweighting and histogramming sampled configurations.
Moreover, the generative model $p_{\theta}(\bm{u} | \bm{s})$ enables direct sampling of the rare-event configurations associated with a given CV.
Thus, a single dataset-free optimization provides both a continuous FES over the CVs and one-shot generation of the corresponding rare-event configurations.

Fig.~\ref{fig:diagram} summarizes both the architecture and the workflow of VaFES.
From bottom to top, one first specifies a target CV value $\bm{s}$; during training, this value is typically sampled from a uniform distribution.
The auxiliary variables, which supply the remaining information required for reversibility, are drawn from a prior distribution, \eg, a Gaussian distribution, and then transformed by a multi-coupling cubic-spline flow $\mathcal{M}_{\theta}$ conditioned on $\bm{s}$ to generate samples $\bm{u}$ from the parameterized model $p_{\theta}(\bm{u} | \bm{s})$.
Together, $\bm{s}$ and $\bm{u}$ define the complete intermediate representation, from which the physical configuration $\bm{x}$ is reconstructed through the inverse bijective CV transformation $\mathcal{T}^{-1}$.
At this stage, both the configuration $\bm{x}$ and the probability density $p_{\theta}(\bm{u} | \bm{s})$ are available for optimization and for FES evaluation through Eq.~(\ref{eq:loss}).
The resulting FES is a continuous, differentiable landscape, enabling methods such as the NEB~\cite{neb1, neb2} and string method~\cite{stringMethod} to identify transition pathways.
When $\bm{s}$ is chosen within the identified rare-event regions, the reconstructed configuration $\bm{x}$ provides a one-shot sample of the corresponding rare event.

\subsection{Application: bistable dimer potential}\label{sec:dimmer}
\begin{figure*}[!htbp]
\centering
\includegraphics[width=1\textwidth]{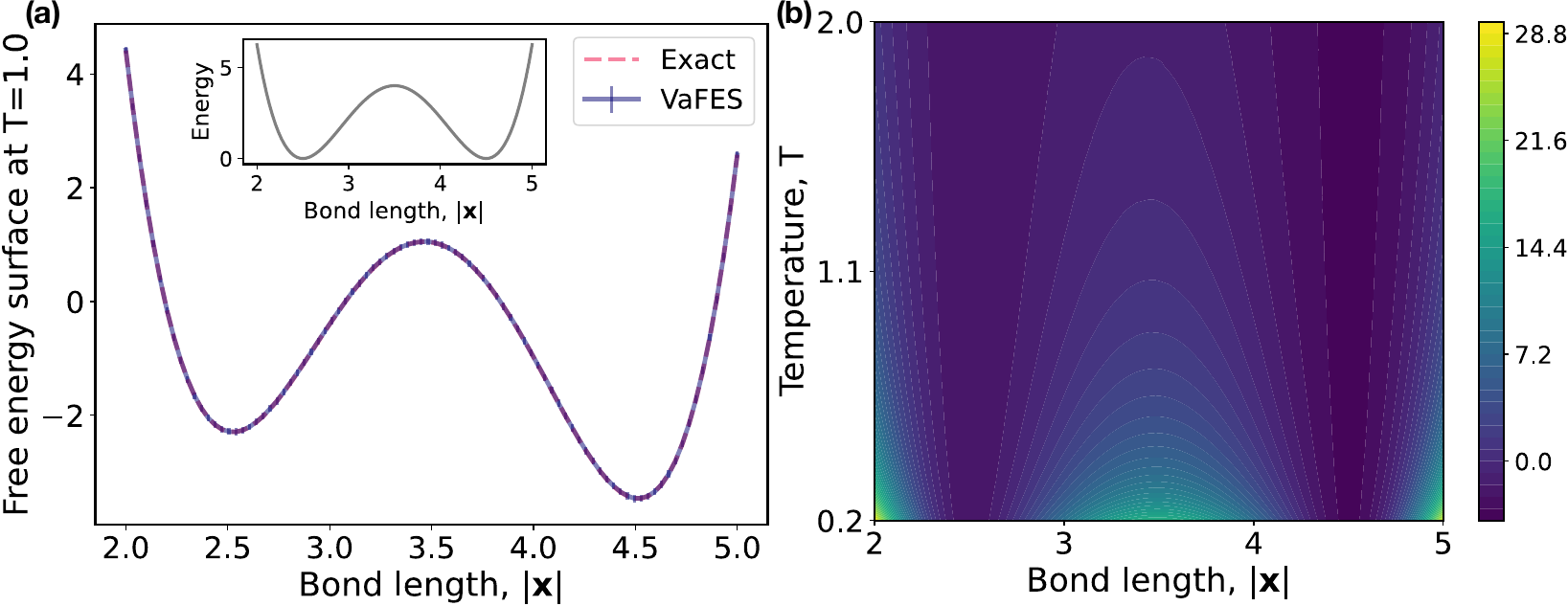}
\caption{
  Free energy surface of the bistable dimer.
  (a) The results obtained by VaFES are shown by the purple solid line and the exact results by the red dashed line. The temperature is fixed at $T=1.0$.
  The inset shows the potential energy landscape of the dimer. The CV is the bond length between the two particles.
  (b) The free energy landscape as a continuous function of temperature and bond length.
}
\label{fig:dimer}
\end{figure*}

As a proof of concept, we first apply VaFES to a simple system for which analytical results are available: the bistable dimer potential, consisting of two particles~\cite{toyDimer}.
The system's energy function depends solely on the distance between the two particles, as in
\begin{equation}
  E(\bm{x}) = 4 (1 - (|\bm{x}| - 3.5)^{2})^2,
  \label{eq:dimerH}
\end{equation}
where $\bm{x}$ denotes the $3$-dimensional vector $(x_0,\ x_1,\ x_2)$ connecting the two particles.
Its magnitude $|\bm{x}|$ is the interparticle distance, which we take as the bond-length CV.
As indicated by Eq.~(\ref{eq:dimerH}) and by the energy landscape shown in the inset of Fig.~\ref{fig:dimer}(a), the potential has two equivalent low-energy valleys at bond lengths of $2.5$ and $4.5$.
Once the entropic contribution to the free energy is included, however, the lower free energy minimum shifts to the more extended state at bond length $4.5$.
This indicates that at finite temperature the system favors an extended bond.

The coarse-grained CV mapping $|\bm{x}| = \sqrt{x_0^2+x_1^2+x_2^2}$ admits the reversible equivalent
\begin{equation}
  (|\bm{x}|,\ u_0,\ u_1) = (\sqrt{x_0^2+x_1^2+x_2^2},\ \sign(x_1)\frac{x_1^2}{x_0^2 + x_1^2},\ \sign(x_2)\frac{x_2^2}{x_0^2 + x_1^2 + x_2^2}),
\end{equation}
where $|\bm{x}|$ is the CV and $u_0$ and $u_1$ are auxiliary variables.
The corresponding energy function in the intermediate representation then follows from Eq.~(\ref{eq:CVE}).
We next optimized VaFES using Eq.~(\ref{eq:loss}), yielding the FES estimates shown in Fig.~\ref{fig:dimer}(a).
Because this is a low-dimensional model, the FES can also be computed analytically by exact integration.
Fig.~\ref{fig:dimer}(a) shows that the VaFES results are in excellent agreement with the exact ones, indicating that VaFES accurately recovers the free energy profile.
The result also confirms the preceding analysis: although the energy has a minimum at bond length $2.5$, the free energy favors the more extended state at bond length $4.5$.

As a further demonstration, we also included temperature $T$ as an external parameter during optimization, allowing VaFES to capture the FES as a continuous, differentiable function of both temperature and bond length.
We plotted these results in Fig.~\ref{fig:dimer}(b).
At low temperatures, the FES gradually approaches the underlying energy landscape, confirming that the preference for the extended bond length is entropic in origin.

\subsection{Application: two paths in diazene evolution}\label{sec:h2n2}

Diazenes constitute an important class of compounds with broad applications in synthesis and materials science.
They are generally planar and can undergo cis--trans isomerization.
The simplest member of this family, \ce{N2H2}, has therefore attracted extensive experimental and theoretical attention.
Two principal mechanisms have been proposed for its cis--trans interconversion: an in-plane inversion pathway and an out-of-plane torsion pathway.
Because the associated activation barriers are similar, the relative contributions and competition between these mechanisms have been the subject of continued discussion~\cite{diazene1, diazene2}.

\begin{figure*}[!htbp]
\centering
\includegraphics[width=1\textwidth]{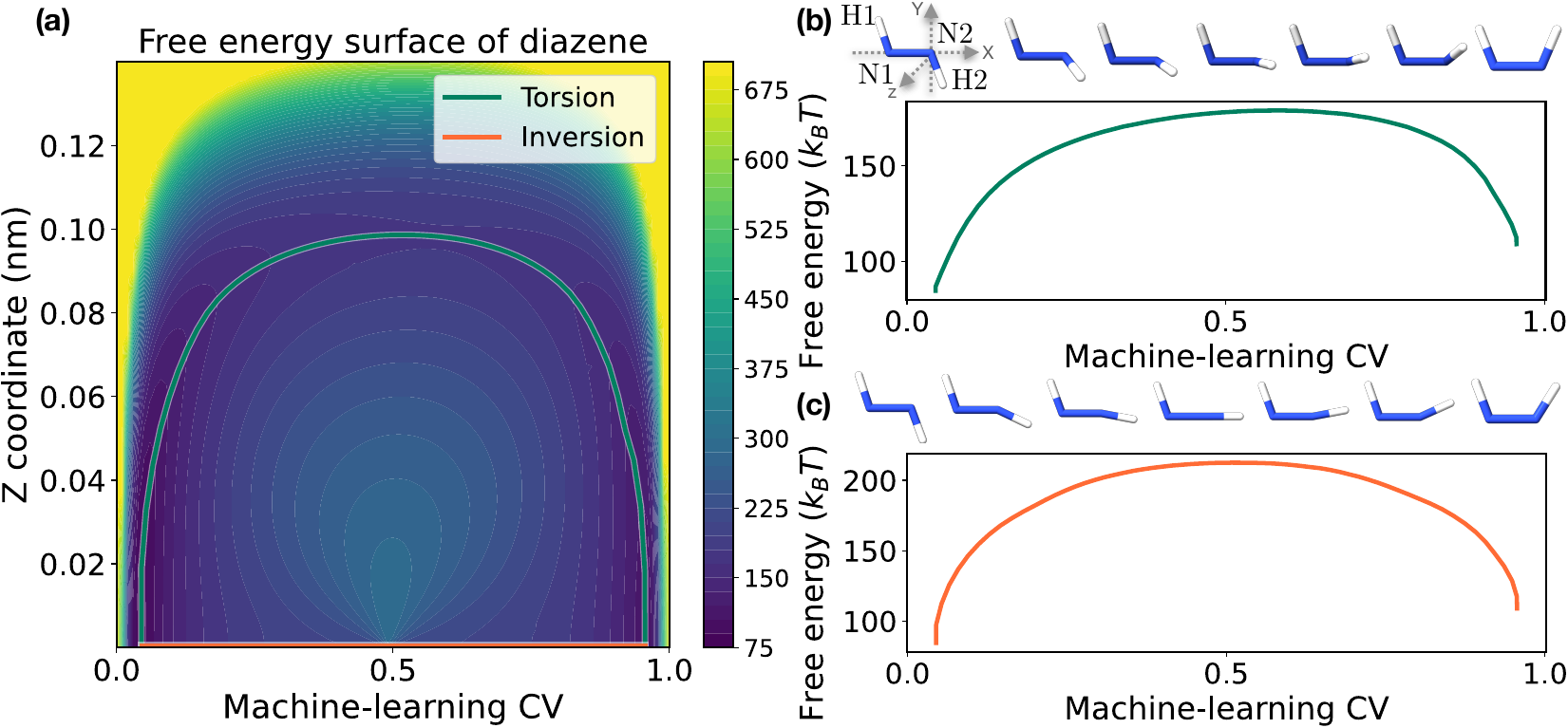}
\caption{
  Free energies of diazene (\ce{N2H2}) at $120$ K.
  (a) Estimated free energy surface of diazene. The CVs are a machine-learning coordinate indicating the trans state at $0$ and the cis state at $1$ on the $x$-axis, and the $z$-displacement of H2 on the $y$-axis. Two reaction paths identified by NEB are shown: the inversion path (red line) and the torsion path (green line).
  (b) Free energy profile along the inversion pathway projected onto the $x$-axis, together with representative diazene configurations. The local-frame coordinate definitions are illustrated using the first configuration.
  (c) Free energy profile along the torsion pathway projected onto the $x$-axis, together with representative diazene configurations.
}
\label{fig:h2n2}
\end{figure*}

To apply VaFES, we first constructed an intermediate representation.
In physical Cartesian space the molecule has $12$ degrees of freedom, of which $6$ can be removed by eliminating global translation and rotation.
Specifically, we fixed atom N2 at the origin, placed atom N1 along the negative $x$-axis, and constrained atom H1 to the $xy$-plane, as shown in  Fig.~\ref{fig:h2n2}(b).
This gives a $6$-dimensional coordinate $(x_1,\ y_1,\ d,\ x_2,\ y_2,\ z_2)$ that retains all necessary information, where $(x_1, y_1)$ is the in-plane position of H1 relative to N1, $d$ is the \ce{N\bond{-}N} bond length, and $(x_2,\ y_2,\ z_2)$ is the full position of H2.

Among these six variables, the $y$- and $z$-coordinates of H2, $(y_2,\, z_2)$, carry the essential information about the isomeric state:
$y_2$ implies the cis and trans configurations, while $z_2$ measures the out-of-plane displacement characterizing the torsion pathway.
Furthermore, because $y_2$ is a geometric coordinate rather than an order parameter that cleanly separates the two states, we introduced a data-driven refinement by training a learnable bijective transformation on $y_2$ to produce a machine-learning CV $s_1$.
Specifically, we parameterized a sigmoid-based transformation $s_1 = L \sigma(k (y_2 - t)) + b$, where $k$ and $t$ are outputs of a multilayer perceptron (MLP) that takes $(x_1,\ y_1,\ d,\ x_2,\ z_2)$ as input. The coefficients $L$ and $b$ provide an affine rescaling that maps the output to $[0, 1]$, with $0$ corresponding to trans and $1$ to cis.
This transformation was trained with a supervised mean-squared-error loss on MD-generated cis and trans configurations.
We therefore selected these two variables as the CVs, \ie, $\bm{s}=(s_1,\, z_2)$, leaving the remaining four variables $(x_1, y_1, d, x_2)$ as auxiliary variables.
This defines an end-to-end bijective transformation from the physical configuration space of diazene to the latent space of the intermediate representation, with an analytically tractable Jacobian.

The resulting FES estimate, obtained by optimizing VaFES with Eq.~(\ref{eq:loss}), is shown in Fig.~\ref{fig:h2n2}(a) and reveals two candidate pathways corresponding to inversion and torsion.
Because VaFES produces a continuous and differentiable FES as a function of the CVs, we can apply the gradient-based NEB~\cite{neb1, neb2} or the string method~\cite{stringMethod} to identify reaction paths with minimum free energy.
Using two different initializations, NEB converges to two smooth pathways, shown in Fig.~\ref{fig:h2n2}(a) as the inversion and torsion paths.
The free energy profile of the inversion pathway is shown in Fig.~\ref{fig:h2n2}(b), and that of the torsion pathway in Fig.~\ref{fig:h2n2}(c).
VaFES also enables direct sampling of physical configurations by specifying CV values along the pathways and completing them with auxiliary variables generated by the trained model.
Representative molecular configurations generated in this way are shown in Fig.~\ref{fig:h2n2}(b,c).
These results are in excellent agreement with previous studies.

This example shows that, in addition to conventional CVs with exact analytical forms, VaFES can also incorporate learnable machine-learning models to define CVs.
Although reversibility is required, there are at least two practical ways to satisfy this condition.
First, as shown here, one can use an intrinsically reversible neural-network structure.
In the present proof of concept, we adopted a simple sigmoid-based neural-network transformation as the machine-learning CV;
however, far more expressive reversible deep-learning models have been developed, \eg, Refs. \cite{inverResNet, ffjord, neuralode, glow}.
Second, many widely used statistical learning methods, such as principal component analysis (PCA) and time-lagged independent component analysis (TICA)~\cite{tica1, tica2}, construct CVs through full-rank linear transformations, which are inherently bijective.
In our next application, we therefore applied TICA to obtain a set of data-driven CV coordinates.

\subsection{Application: alanine dipeptide using data-driven CVs}\label{sec:aldp}

As a third example, we consider transitions in alanine dipeptide in vacuum, where the rare-event configurations are more intricate and physically relevant.
Unlike the diazene case, which used a combination of one analytical CV and one machine-learning CV, here we adopt a fully data-driven set of CVs obtained from TICA.

\begin{figure*}[!htbp]
\centering
\includegraphics[width=1\textwidth]{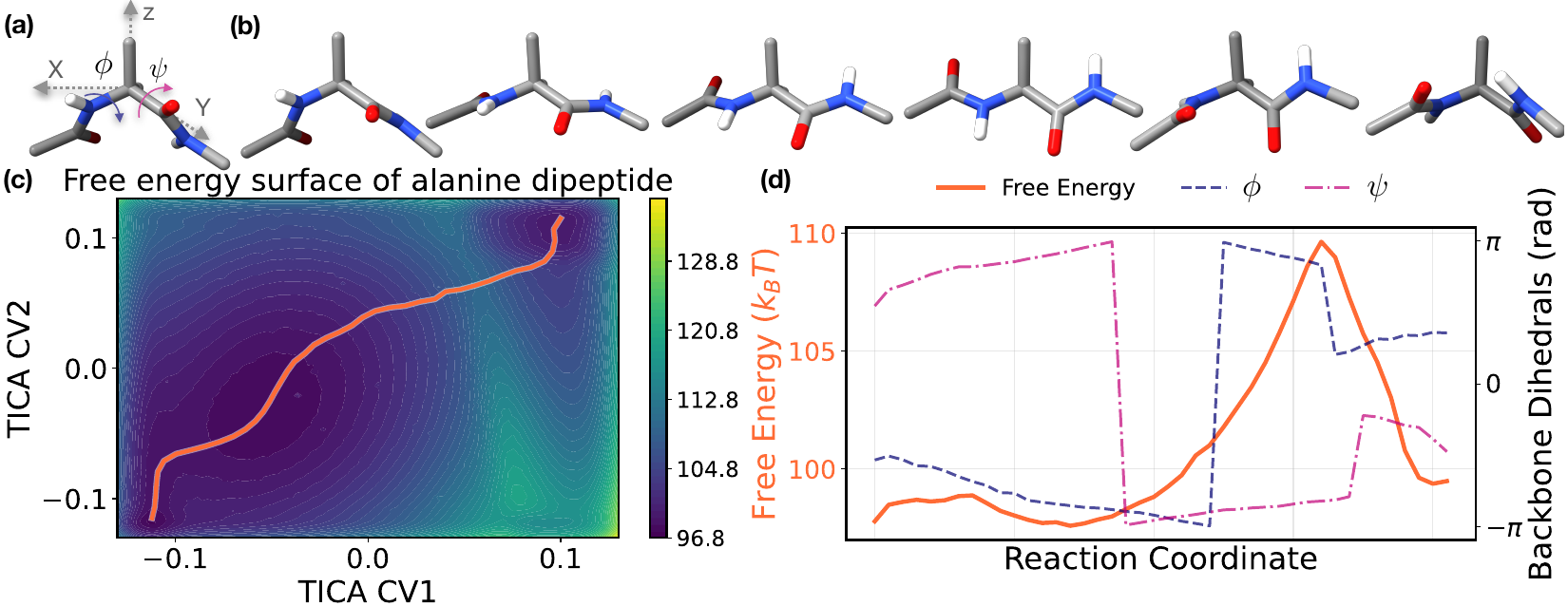}
\caption{
    Structural change and free energy landscape of alanine dipeptide at 300 K.
    (a) The structure of the dipeptide molecule defines the local-frame coordinate and the two backbone dihedral angles $\psi$ and $\phi$.
(b) Molecular configurations along the reaction path identified by the NEB method.
  (c) Estimated free energy surface of alanine dipeptide obtained by VaFES. The CV coordinates are the two dominant components identified by TICA. A reaction path (red line) was identified using NEB, with endpoints fixed at the two local minima separated by the greatest distance.
  (d) Free energy profile along the reaction path in (c).
  The backbone dihedral angles, $\psi$ and $\phi$, are shown by the dashed blue and dash-dot pink lines, respectively.   
}
\label{fig:peptide}
\end{figure*}

For a concise representation, we first treated all methyl groups as rigid bodies.
In physical Cartesian space, alanine dipeptide thus comprises $13$ atoms, yielding $39$ degrees of freedom.
We then eliminated six more dimensions by fixing translation and rotation through a constrained coordinate system, as in the diazene example.
Specifically, we fixed the \ce{C_\alpha} atom at the origin, placed the C-terminal carbonyl carbon along the positive $y$-axis, constrained the amide nitrogen attached to \ce{C_\alpha} to the $xy$-plane, and pointed the \ce{C_{\beta}} carbon to the positive $z$-axis, resulting in a $33$-dimensional coordinate that retains all internal degrees of freedom.
We show the coordinate system in Fig.~\ref{fig:peptide}(a).

To define the CVs, we performed TICA on configuration data generated from long MD simulations.
The MD trajectory was first converted into the $33$-dimensional coordinate representation, after which TICA was used to identify the dominant slow modes.
The first two components, which capture the most relevant collective behaviors, were selected as the CVs.
Notably, the TICA projection from the $33$-dimensional coordinate to independent-component space, after whitening, is an orthogonal transformation, and is therefore bijective, satisfying the requirement of VaFES.
An additional benefit of this orthogonal transformation is that the determinant of its Jacobian is simply $1$, giving a particularly concise form of Eq.~(\ref{eq:CVE}).
Aside from the two most prominent components, all remainings were treated as auxiliary variables.

Based on this intermediate representation, we trained the cubic-spline flow following the VaFES procedure described in Sec.~\ref{sec:vafes}.
The resulting FES is shown in Fig.~\ref{fig:peptide}(c).
Because the VaFES estimate is continuous and differentiable with respect to the CVs, the NEB method can again be used to identify a minimum-free-energy reaction path.
We selected the two local minima separated by the largest distance in CV space as two NEB endpoints, and obtained the path shown in Fig.~\ref{fig:peptide}(c).
The free energy profile along this path is shown in Fig.~\ref{fig:peptide}(d). 
It is well known that the two backbone dihedral angles $\psi$ and $\phi$ defined in Fig.~\ref{fig:peptide}(a) dominate the collective behavior of alanine dipeptide.
We therefore also computed the corresponding values of $\psi$ and $\phi$ along the reaction path, as shown in Fig.~\ref{fig:peptide}(d).
Our result is in excellent agreement with previous ones.
First, the two TICA CVs effectively capture the two backbone dihedral angles, as different CV locations along the reaction path correspond to different $(\psi,\phi)$ configurations.
Second, the low- and high-free-energy regions, together with their associated dihedral-angle values, agree well with the widely accepted free energy landscape in the dihedral-angle space.
By complementing the CV values with auxiliary variables generated by the VaFES cubic-spline flow, we recovered the full molecular configurations.
Figure~\ref{fig:peptide}(b) shows some key transition structures along the NEB path.

\subsection{Application: chignolin}\label{sec:chignolin}

After successfully validating our method using three well-studied examples, we now turn to a more complicated case.
We consider the $10$-residue protein chignolin~\cite{chignolin}, which may fold into a stable $\beta$-hairpin.
To challenge the VaFES method, we chose the native structure (PDB 1UAO), which has multiple metastable states as the misfolded and intermediate structures, and is thus more difficult to study than its more stable mutant CLN025~\cite{1uao, 1uao2}. 
We modeled the system using the \texttt{Amber14} force field and the \texttt{GBN2} implicit water model at $350$\,K.

To construct a concise representation of the protein, we ignored all hydrogen atoms from explicit sampling, 
which can be easily recovered during energy evaluation using local molecular geometry~\cite{addH}, thereby reducing the chignolin molecule to $77$ heavy atoms.
We further reduced six global translational and rotational degrees of freedom by fixing the molecule in a restricted frame, putting \ce{C_{\alpha}^8} at the origin, \ce{C_\alpha^3} on the positive $y$-axis, and \ce{C_\alpha^1} in the $x$-$y$ plane.
After these, there are $225$ internal degrees of freedom left.
As for CVs, we selected two inter-\ce{C_\alpha} distances that monitor folding progress: $|\ce{C_\alpha^3\bond{-}C_\alpha^8}|$, which captures the key hydrogen-bond contact across the $\beta$-hairpin turn~\cite{CA3CA8}, and $|\ce{C_\alpha^1\bond{-}C_\alpha^{10}}|$, which measures end-to-end compaction.
These CVs form the first two dimensions of our intermediate representation, while the remaining $223$ auxiliary dimensions are optimized by VaFES via a cubic-spline flow.

Unlike the small molecules in previous examples, proteins require a highly structured bijective transformation to convert Cartesian coordinates into an intermediate representation suitable for variational modeling.
Including the CV dimensions, this process maps all $225$ dimensions into physically meaningful internal coordinates, such as bond lengths, bond angles, and dihedral angles.
We therefore constructed a composite bijection by sequentially applying four elementary reversible maps, each with a closed-form Jacobian determinant, to specific heavy-atom subsets according to their chemical functions:
(1) a spherical coordinate transformation that converts displacement vectors between these CV-related \ce{C_\alpha} atoms into the two distance CVs plus auxiliaries of angular and distance variables;
(2) a local-frame coordinate transformation that, for each residue, decomposes its backbone triad (\ce{N\bond{-}C_\alpha\bond{-}C}) into Euler angles and frame-geometry scalars,
and represents all its atoms in the triad-defined local frame.
(3) a torsion transformation representing terminal atoms in bonded quadruples via bond lengths, bond angles, and dihedral angles; and
(4) a planar local-frame transformation expressing aromatic rings and carboxylate groups in local 2D frames.
Consequently, the full $225$-dimensional intermediate representation comprises the $2$ CVs, $4$ angular and distance variables, $30$ Euler angles, $41$ dihedral angles, $41$ bond lengths, $41$ bond angles, and $66$ local-frame variables.
Full mathematical definitions and atom-level assignments are provided in the Supplementary Information~\cite{SM}.

To model this high-dimensional and complex system, we replaced the simple MLPs used in the previous examples with a more expressive neural-network architecture.
Specifically, the spline parameters in each coupling layer are generated by a hybrid residual network (ResNet)~\cite{resnet} and MLP architecture, in which the inputs are first projected onto high-dimensional feature vectors by an MLP and then processed by residual one-dimensional convolutional blocks.
Finally, to reduce the risk of trapping in local minima, we incorporated temperature as an additional external parameter using the variational temperature-differentiable (VaTD) framework developed by us previously~\cite{vatd}.

\begin{figure*}[!htbp]
\centering
\includegraphics[width=1\textwidth]{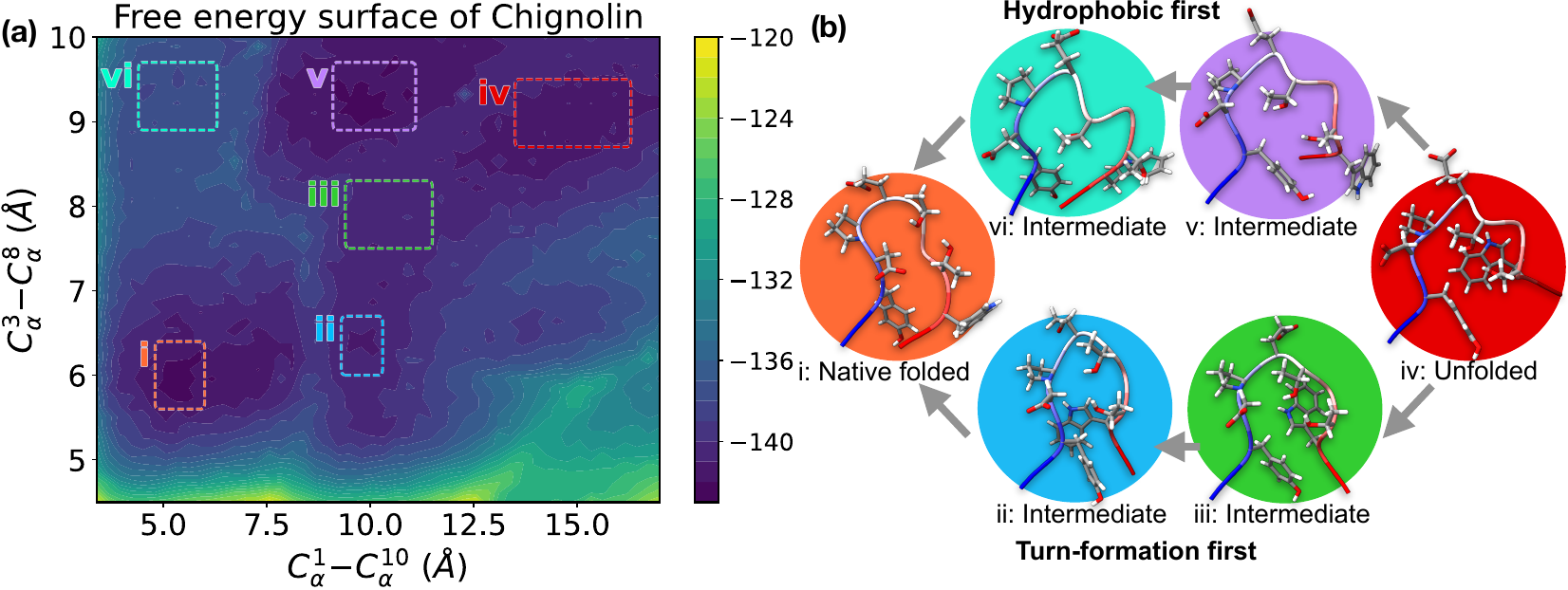}
\caption{
Free energy profile and key configurations of chignolin (PDB 1UAO) at $350$ K.
  (a) The estimated free energy surface. The CVs are the two inter-\ce{C_\alpha} distances $|\ce{C_\alpha^1\bond{-}C_\alpha^{10}}|$ and $|\ce{C_\alpha^3\bond{-}C_\alpha^{8}}|$. Six regions in CV space are boxed and labeled as key states.
  (b) Two folding pathways implied by the estimated FES. In the upper pathway, folding is initiated by 
the hydrophobic collapse from unfolded state iv to intermediate state v then vi, followed by turn formation from intermediate state vi to native folded state i.  
In the lower pathway, folding is initiated by turn formation: the two $\beta$-strands approach each other with modest hydrophobic closure and nascent turn formation from unfolded state iv to intermediate state iii, followed by tight turn formation from intermediate state iii to intermediate state ii and final hydrophobic collapse from intermediate state ii to native folded state i.
}
\label{fig:chignolin}
\end{figure*}

The FES obtained by our VaFES method suggests two possible folding pathways as shown in Fig.~\ref{fig:chignolin}.
The upper pathway in Fig.~\ref{fig:chignolin}(b) begins with hydrophobic closure between \ce{C_\alpha^1} and \ce{C_\alpha^10}, followed by turn formation that reduces the distance between \ce{C_\alpha^3} and \ce{C_\alpha^8}.
This pathway has a relatively high free energy barrier, suggesting that it is unlikely to be the dominant folding route of chignolin.
The lower pathway in Fig.~\ref{fig:chignolin}(b), which is more likely, has a smaller barrier and proceeds in three stages.
Folding first involves hydrophobic assembly of the two aromatic rings together with the formation of a nascent turn, leading to a moderate simultaneous reduction in both $|\ce{C_\alpha^1\bond{-}C_\alpha^{10}}|$ and $|\ce{C_\alpha^3\bond{-}C_\alpha^8}|$ as the system evolves from the unfolded state iv to intermediate state iii.
Next, $|\ce{C_\alpha^3\bond{-}C_\alpha^8}|$ decreases from more than 8 \AA\ to  about 6 \AA\ with almost unchanged $|\ce{C_\alpha^1\bond{-}C_\alpha^{10}}|$, indicating formation of a tight turn structure in the transition from intermediate state iii to intermediate state ii.
Final folding then proceeds with a further collapse of the hydrophobic core, transforming intermediate state ii into the native folded state i.
These findings agree well with long-timescale MD simulations~\cite{folding2}, the FES reported in a previous study~\cite{folding1}, and general views~\cite{folding3, folding4} suggesting that local structure forms early, followed by later specific core packing during protein folding.
As this is primarily a methodology study, we do not claim a definitive folding mechanism for chignolin, because the detailed folding behavior also heavily depends on the choice of force field and on whether the solvent is modeled explicitly or implicitly~\cite{1uao}. Instead, our goal is to test the technical capabilities of the VaFES framework.

At the microscopic level, the native folded state i is identified as the local minimum in the lower-right region of the FES, where both CVs take small values.
These configurations exhibited compact $\pi$-turns with radii of about $5$--$6\si{\angstrom}$, consistent with previous MD results~\cite{folding1}.
Compared to the experimental NMR structure, the directly sampled native folded structures achieved a minimum \ce{C_{\alpha}}-RMSD of about $1\si{\angstrom}$ and remained within $1.5\si{\angstrom}$, comparable to the level reported in Ref.~\cite{chignolin}.
These results show that VaFES captures the dominant microscopic folding structure of chignolin with high fidelity, further implying the accuracy of the estimated FES.
More broadly, these also highlight the ability of VaFES to recover one-shot realistic rare-event configurations directly from the energy function, without pre-generated datasets or extended MCMC/MD simulations.

\section{Conclusion}
In this paper, we introduced VaFES, a dataset-free variational framework that uses tractable-density generative models to compute the FES directly from the system energy function.
This enables subsequent identification of rare-event pathways with methods such as the NEB~\cite{neb1, neb2} and the string method~\cite{stringMethod}, as well as one-shot generation of rare-event configurations by direct sampling from the generative model.
Compared with conventional sampling-based approaches~\cite{umbrellaSampling, TI1, TI2, metaD, WangLandau}, which rely on costly rare-event sampling through MCMC or MD, VaFES instead follows an FES-first strategy that targets the free energy landscape directly.
Compared with existing deep-learning variational methods~\cite{prl, prr, prx, wuDianPRL, vatd, atomicNF1, atomicNF2, LWnewPRL}, which require the exact energy in the original full configuration space and therefore break down after coarse-graining onto CVs, VaFES overcomes this obstacle by replacing the coarse-grained CV transformation with a reversible equivalent.
By the universality of bijective transformations, any coarse-grained transformation can be rigorously expanded into a bijection by incorporating auxiliary variables, in a process that can be automated by compiler-like algorithms~\cite{al2012reversible}.
This construction admits a latent space of intermediate representation, a joint of coarse-grained CV and auxiliaries, with the same dimensionality as the original full space.
The reversibility further allows the energy to be rewritten exactly in this latent space, making the FES directly accessible to variational optimization.

This proposed framework offers several advantages.
As a variational optimization scheme, the deep-learning model is trained on self-generated one-shot configurations, without the need for pre-generated datasets.
In contrast to the histogrammed, discrete free energy profiles produced by conventional deep-learning approaches~\cite{frankNoe, conditionalBG, conditionedBGtps, enhancedDiffusion}, VaFES yields the FES as a continuous, differentiable function of the CVs, enabling gradient-based methods such as the NEB and string method for the identification of rare events.
Rare-event configurations can then be sampled directly through one-shot generation from the trained model conditioned on the corresponding CV values.
Thus, a single variational optimization yields both a continuous, differentiable FES and direct one-shot generation of rare-event configurations.
Furthermore, VaFES constructs its latent space by expanding coarse-grained CV transformations into bijections.
The resulting latent variables, the intermediate representation, remain physically interpretable by construction, as the CVs explicitly occupy a subset of its dimensions, thereby preserving compatibility with established CV-based intuition.
By the universality of bijections, this construction accommodates arbitrary CV formulations with no restriction, including data-driven coordinates, machine-learned and statistically derived CVs, and advanced deep-learning reaction coordinates~\cite{aicv1, aicv2, aicv3}.
The intermediate representation also provides a practical route to handling symmetry and geometric constraints in complex systems.
In contrast to approaches that encode physical structure directly into specialized equivariant architectures~\cite{equivariantFlowNoe, klein2023equivariant, equivariantFlowLattice}, VaFES offers an alternative by transforming the system into coordinates in which global symmetries are removed systematically and local constraints are imposed exactly.
More broadly, VaFES is compatible with any tractable-density generative model and can in principle be combined with more accurate energy functions, including quantum Hamiltonians.
By linking rare-event configuration generation, coarse-grained thermodynamics, and a differentiable free energy landscape within a single framework, VaFES offers a systematic and scalable route to studying transition behavior and rare events in complex statistical systems.

\section{Methods}

\subsection{Variational training of tractable-density models over a continuous external parameter}\label{sec:model}

Generative deep-learning models are designed to produce samples from complex high-dimensional distributions, such as natural images~\cite{generativeModel,deeplearningBook, glow}.
Tractable-density generative models form a special class for which the probability density is explicitly available.
They can therefore both evaluate exact sample probability and draw one-shot samples directly from the learned distribution.
Representative examples include autoregressive models, such as PixelCNN and PixelRNN~\cite{pixelRCNN}, and normalizing flows (NFs), such as RealNVP~\cite{realnvp} and the cubic-spline flow.

Variational training of a tractable-density model starts from a physical system with an explicit energy function $E(\bm{x})$, where $\bm{x}$ denotes a configuration.
The objective is to minimize the KLD between the model distribution $p_{\theta}(\bm{x})$ and the Boltzmann distribution $\pi(\bm{x})$ defined by $E$:
\begin{equation}
  \begin{split}
  D_{KL}\big(p_{\theta}(\bm{x})|| \pi(\bm{x})\big) &= \underset{\bm{x}\sim p_{\theta}}{\mathbb{E}} \left[\ln p_{\theta}(\bm{x}) - \ln \pi(\bm{x})\right] \\
  &= \ln \mathcal{Z} + \underset{\bm{x}\sim p_{\theta}}{\mathbb{E}} \left[\ln p_{\theta}(\bm{x}) + \beta E(\bm{x})\right].
    \label{eq:kld2}
    \end{split}
\end{equation}
Because the KLD is non-negative, \ie,
$\underset{\bm{x}\sim p_{\theta}}{\mathbb{E}} [\ln p_{\theta}(\bm{x}) + \beta E(\bm{x})] \ge -\ln \mathcal{Z}$,
the estimator $\underset{\bm{x}\sim p_{\theta}}{\mathbb{E}} [\ln p_{\theta}(\bm{x}) + \beta E(\bm{x})]$ provides an upper bound on the exact free energy $-\ln \mathcal{Z}$.
Minimizing the KLD therefore yields a variational free energy estimate.
At the optimum, the model distribution approaches the target Boltzmann distribution; because the model is generative, physical configurations can then be sampled directly in a one-shot style.

Reference~\cite{vatd} introduced the VaTD framework, which extends this optimization to a continuously varying external parameter such as temperature.
In that setting, the integral form of Eq.~(\ref{eq:kld2}) over a continuous temperature range is estimated using randomly sampled temperatures:
\begin{equation}
  \underset{\theta}{\text{min}} \underset{\beta \sim U, \bm{x}\sim p_{\theta}}{\mathbb{E}} \left[\ln p_{\theta}(\bm{x};\, \beta) + \beta E(\bm{x}) \right]
  \label{eq:loss2}
\end{equation}
After optimization, the free energy estimate becomes a continuous, differentiable function of temperature.
The trained model also acquires the ability to generate accurate samples across a continuous range of temperature.
The VaTD formulation is not restricted to temperature and can be generalized to other external parameters of the system.

\subsection{Cubic-spline flow}

A NF~\cite{normalizingflow1, normalizingflow2} models a complex distribution as a composition of bijective transformations.
Starting from latent variables $\bm{z}$ drawn from a simple prior $q$ (\eg, uniform or Gaussian), the flow applies a parameterized bijection $\bm{x} = \mathcal{M}_{\theta}(\bm{z})$ to generate samples from the target distribution.
Because the mapping is bijective, the probability density of $\bm{x}$ is obtained exactly from the change-of-variables formula:
\begin{equation}
  \ln p_{\theta}(\bm{x}) = \ln q(\bm{z}) - \ln \left|\det\frac{\partial \mathcal{M}_{\theta}}{\partial \bm{z}}\right|.
  \label{eq:distNF}
\end{equation}
In practice, $\mathcal{M}_{\theta}$ is implemented as a stack of coupling layers~\cite{realnvp}.
Each layer partitions the variables into two groups: one remains unchanged and conditions a neural network whose outputs parameterize an elementwise reversible transformation of the other group.
This construction yields a triangular Jacobian, so the determinant reduces to a product of one-dimensional derivatives and can be evaluated efficiently.

In a cubic-spline flow~\cite{cubic}, each coupling layer uses an elementwise monotonic piecewise-cubic spline.
For $K$ bins on $[0,1]$, the parameter network outputs $K$ bin widths, $K$ bin heights and two boundary derivatives for each transformed variable.
These quantities define a continuously differentiable monotone cubic spline through Steffen interpolation~\cite{fritsch1980monotone, steffen1990simple}.
As $K$ increases, the spline can approximate any continuous monotone function arbitrarily well while retaining an analytically tractable Jacobian determinant.

\subsection{Universality of bijective transformation}\label{sec:bijection}

A fundamental result in reversible computing is the universality of bijective transformations: for any deterministic map $\bm{s} = \mathcal{F}(\bm{x})$, there exists an equivalent bijective program $\mathcal{T}$ acting on an extended state space, $(\bm{x}, \bm{u}') \rightarrow (\bm{s}, \bm{u})$.
Here $\bm{u}'$ denotes auxiliary bits or variables, typically initialized to zero, and $\bm{u}$ denotes the transformed anciliaries.
Crucially, $\mathcal{T}$ preserves all input information, so it can be reversed exactly while still returning $\bm{s} = \mathcal{F}(\bm{x})$ on the primary output~\cite{reversibleComp2, reversibleCompBook}.

From a thermodynamic perspective, preserving information through auxiliary bits avoids the Landauer cost of $kT\ln 2$ per erased bit and provides a theoretical route to energy-efficient computation.
Combined with universality, this implies that any computation can in principle be implemented without intrinsic dissipation, approaching thermodynamic reversibility at the Landauer limit~\cite{Landauer1961}.
The same principle also underlies quantum computation and quantum circuits~\cite{nielsen2010quantum}, linking reversible classical computation to universal quantum processing.

This paradigm also admits constructive automation: compiler-like procedures can synthesize a reversible program $\mathcal{T}$ directly from $\mathcal{F}$~\cite{al2012reversible}.
Such synthesis is essential in practical quantum computing, where classical functions must be converted into unitary gate sequences while preserving reversibility~\cite{saeedi2013synthesis}.

\nocite{SM}

\section{Code availability}
Our code implementation is opensourced on Github~\cite{Code}.

\begin{acknowledgments}
We thank Tao Li and Chu Li for many useful discussions.
S.-H.L. and D.P. acknowledge support from Hong Kong Research Grants Council (RGC) (GRF-16302423).
D.P. acknowledges support from Hong Kong RGC (GRF-16301723, GRF-16310225), 
National Natural Science Foundation of China/RGC Joint Research Scheme (N\_HKUST664/24). 
Part of this work was carried out using computational resources from the National Supercomputer Center in Guangzhou, China, and the X-GPU cluster supported by the RGC Collaborative Research Fund C6021-19EF.
\end{acknowledgments}

\bibliography{ref}

@misc{SM,
note={
See the SI for details about
(a). the force field information, the deep-learning model structure, and extra parameters used in each applications;
(b) the detailed scheme of bijective transformations used in each applications.
The SI cites ~\cite{resnet, cubic, vatd, 54A7, atb1, atb3}.
}
}

@misc{Code,
  note={See \href{https://github.com/li012589/vafes}{https://github.com/li012589/vafes} for a PyTorch implementation.}
}

@article{vfep,
	author = {Lee, Tai-Sung and Radak, Brian K. and Pabis, Anna and York, Darrin M.},
	date = {2013/01/08},
	doi = {10.1021/ct300703z},
	isbn = {1549-9618},
	journal = {Journal of Chemical Theory and Computation},
	journal1 = {Journal of Chemical Theory and Computation},
	journal2 = {J. Chem. Theory Comput.},
	month = {01},
	number = {1},
	pages = {153--164},
	publisher = {American Chemical Society},
	title = {A New Maximum Likelihood Approach for Free Energy Profile Construction from Molecular Simulations},
	type = {doi: 10.1021/ct300703z},
	url = {https://doi.org/10.1021/ct300703z},
	volume = {9},
	year = {2013},
	year1 = {2013}}

@article{PRLvfe,
  title = {Variational Approach to Enhanced Sampling and Free Energy Calculations},
  author = {Valsson, Omar and Parrinello, Michele},
  journal = {Phys. Rev. Lett.},
  volume = {113},
  issue = {9},
  pages = {090601},
  numpages = {5},
  year = {2014},
  month = {Aug},
  publisher = {American Physical Society},
  doi = {10.1103/PhysRevLett.113.090601},
  url = {https://link.aps.org/doi/10.1103/PhysRevLett.113.090601}
}

@article{TI1,
title = {Constrained reaction coordinate dynamics for the simulation of rare events},
journal = {Chemical Physics Letters},
volume = {156},
number = {5},
pages = {472-477},
year = {1989},
issn = {0009-2614},
doi = {https://doi.org/10.1016/S0009-2614(89)87314-2},
url = {https://www.sciencedirect.com/science/article/pii/S0009261489873142},
author = {E.A. Carter and Giovanni Ciccotti and James T. Hynes and Raymond Kapral}
}

@article{TI2,
    author = {Sprik, Michiel and Ciccotti, Giovanni},
    title = {Free energy from constrained molecular dynamics},
    journal = {The Journal of Chemical Physics},
    volume = {109},
    number = {18},
    pages = {7737-7744},
    year = {1998},
    month = {11},
    issn = {0021-9606},
    doi = {10.1063/1.477419},
    url = {https://doi.org/10.1063/1.477419},
    eprint = {https://pubs.aip.org/aip/jcp/article-pdf/109/18/7737/19114804/7737\_1\_online.pdf},
}

@article{umbrellaSampling,
title = {Nonphysical sampling distributions in Monte Carlo free-energy estimation: Umbrella sampling},
journal = {Journal of Computational Physics},
volume = {23},
number = {2},
pages = {187-199},
year = {1977},
issn = {0021-9991},
doi = {https://doi.org/10.1016/0021-9991(77)90121-8},
url = {https://www.sciencedirect.com/science/article/pii/0021999177901218},
author = {G.M. Torrie and J.P. Valleau}
}

@article{WangLandau,
  title = {Efficient, Multiple-Range Random Walk Algorithm to Calculate the Density of States},
  author = {Wang, Fugao and Landau, D. P.},
  journal = {Phys. Rev. Lett.},
  volume = {86},
  issue = {10},
  pages = {2050--2053},
  numpages = {0},
  year = {2001},
  month = {Mar},
  publisher = {American Physical Society},
  doi = {10.1103/PhysRevLett.86.2050},
  url = {https://link.aps.org/doi/10.1103/PhysRevLett.86.2050}
}

@article{metaD,
author = {Alessandro Laio  and Michele Parrinello },
title = {Escaping free-energy minima},
journal = {Proceedings of the National Academy of Sciences},
volume = {99},
number = {20},
pages = {12562-12566},
year = {2002},
doi = {10.1073/pnas.202427399},
URL = {https://www.pnas.org/doi/abs/10.1073/pnas.202427399},
eprint = {https://www.pnas.org/doi/pdf/10.1073/pnas.202427399}}

@article{metaDreview1,
author = {Barducci, Alessandro and Bonomi, Massimiliano and Parrinello, Michele},
title = {Metadynamics},
journal = {WIREs Computational Molecular Science},
volume = {1},
number = {5},
pages = {826-843},
doi = {https://doi.org/10.1002/wcms.31},
url = {https://wires.onlinelibrary.wiley.com/doi/abs/10.1002/wcms.31},
eprint = {https://wires.onlinelibrary.wiley.com/doi/pdf/10.1002/wcms.31},
year = {2011}
}

@article{metaDreview2,
doi = {10.1088/0034-4885/71/12/126601},
url = {https://dx.doi.org/10.1088/0034-4885/71/12/126601},
year = {2008},
month = {nov},
publisher = {},
volume = {71},
number = {12},
pages = {126601},
author = {Laio, Alessandro and Gervasio, Francesco L},
title = {Metadynamics: a method to simulate rare events and reconstruct the free energy in biophysics, chemistry and material science},
journal = {Reports on Progress in Physics}
}

@article{Landauer1961,
  title={Irreversibility and heat generation in the computing process},
  author={Landauer, Rolf},
  journal={IBM journal of research and development},
  volume={5},
  number={3},
  pages={183--191},
  year={1961},
  publisher={Ibm}
}

@article{saeedi2013synthesis,
  title={Synthesis and optimization of reversible circuits—a survey},
  author={Saeedi, Mehdi and Markov, Igor L},
  journal={ACM Computing Surveys (CSUR)},
  volume={45},
  number={2},
  pages={1--34},
  year={2013},
  publisher={ACM New York, NY, USA}
}

@book{al2012reversible,
  title={Reversible logic synthesis: from fundamentals to quantum computing},
  author={Al-Rabadi, Anas N},
  year={2012},
  publisher={Springer Science \& Business Media}
}

@book{nielsen2010quantum,
  title={Quantum computation and quantum information},
  author={Nielsen, Michael A and Chuang, Isaac L},
  year={2010},
  publisher={Cambridge university press}
}

@ARTICLE{reversibleComp2,
  author={Bennett, C. H.},
  journal={IBM Journal of Research and Development}, 
  title={Logical Reversibility of Computation}, 
  year={1973},
  volume={17},
  number={6},
  pages={525-532},
  doi={10.1147/rd.176.0525}}

@book{reversibleCompBook,
  title={Introduction to reversible computing},
  author={Perumalla, Kalyan S},
  year={2013},
  publisher={CRC Press}
}

@misc{generativeModel,
      title={NIPS 2016 Tutorial: Generative Adversarial Networks}, 
      author={Ian Goodfellow},
      year={2017},
      eprint={1701.00160},
      archivePrefix={arXiv},
      primaryClass={cs.LG},
      url={https://arxiv.org/abs/1701.00160}, 
}

@article{prl,
title = {Neural Network Renormalization Group},
  author = {Li, Shuo-Hui and Wang, Lei},
  journal = {Phys. Rev. Lett.},
  volume = {121},
  issue = {26},
  pages = {260601},
  numpages = {7},
  year = {2018},
  month = {Dec},
  publisher = {American Physical Society},
  doi = {10.1103/PhysRevLett.121.260601},
  url = {https://link.aps.org/doi/10.1103/PhysRevLett.121.260601},
archivePrefix = {arXiv},
arxivId = {1802.02840},
eprint = {1802.02840},
url = {http://arxiv.org/abs/1802.02840},
year = {2018}
}

@article{prx,
  title = {Neural Canonical Transformation with Symplectic Flows},
  author = {Li, Shuo-Hui and Dong, Chen-Xiao and Zhang, Linfeng and Wang, Lei},
  journal = {Phys. Rev. X},
  volume = {10},
  issue = {2},
  pages = {021020},
  numpages = {13},
  year = {2020},
  month = {Apr},
  publisher = {American Physical Society},
  doi = {10.1103/PhysRevX.10.021020},
  url = {https://link.aps.org/doi/10.1103/PhysRevX.10.021020},
    archivePrefix={arXiv},
    arxivId = {1910.00024},
    eprint={1910.00024},
    url = {http://arxiv.org/abs/1910.00024}
}

@article{prr,
  title = {Machine learning holographic mapping by neural network renormalization group},
  author = {Hu, Hong-Ye and Li, Shuo-Hui and Wang, Lei and You, Yi-Zhuang},
  journal = {Phys. Rev. Research},
  volume = {2},
  issue = {2},
  pages = {023369},
  numpages = {12},
  year = {2020},
  month = {Jun},
  publisher = {American Physical Society},
  doi = {10.1103/PhysRevResearch.2.023369},
  url = {https://link.aps.org/doi/10.1103/PhysRevResearch.2.023369},
  archivePrefix = {arXiv},
arxivId = {1903.00804},
eprint = {1903.00804},
url = {https://arxiv.org/abs/1903.00804},
year = {2019}

}

@article{vatd,
  title = {Deep generative modeling of the canonical ensemble with differentiable thermal properties},
  author = {Li, Shuo-Hui and Zhang, Yao-Wen and Pan, Ding},
  journal = {Phys. Rev. Lett.},
  pages = {--},
  year = {2025},
  month = {Jun},
  publisher = {American Physical Society},
  doi = {10.1103/8wx7-kyx8},
  url = {https://link.aps.org/doi/10.1103/8wx7-kyx8}
}

@article{toyDimer,
author = {Jerome P. Nilmeier  and Gavin E. Crooks  and David D. L. Minh  and John D. Chodera },
title = {Nonequilibrium candidate Monte Carlo is an efficient tool for equilibrium simulation},
journal = {Proceedings of the National Academy of Sciences},
volume = {108},
number = {45},
pages = {E1009-E1018},
year = {2011},
doi = {10.1073/pnas.1106094108},
URL = {https://www.pnas.org/doi/abs/10.1073/pnas.1106094108},
eprint = {https://www.pnas.org/doi/pdf/10.1073/pnas.1106094108}
}

@article{diazene1,
author = {Hwang, Der-Yan and Mebel, Alexander M.},
title = {Reaction Mechanism of N2/H2 Conversion to NH3: A Theoretical Study},
journal = {The Journal of Physical Chemistry A},
volume = {107},
number = {16},
pages = {2865-2874},
year = {2003},
doi = {10.1021/jp0270349},
URL = {https://doi.org/10.1021/jp0270349},
eprint = {https://doi.org/10.1021/jp0270349}
}

@article{diazene2,
    author = {Sand, Andrew M. and Schwerdtfeger, Christine A. and Mazziotti, David A.},
    title = {Strongly correlated barriers to rotation from parametric two-electron reduced-density-matrix methods in application to the isomerization of diazene},
    journal = {The Journal of Chemical Physics},
    volume = {136},
    number = {3},
    pages = {034112},
    year = {2012},
    month = {01},
    issn = {0021-9606},
    doi = {10.1063/1.3675683},
    url = {https://doi.org/10.1063/1.3675683},
}

@article{chatGPT,
  title={Improving Language Understanding by Generative Pre-Training},
  author={Radford, Alec and Narasimhan, Karthik and Salimans, Tim and Sutskever, Ilya},
  journal={OpenAI Blog},
  volume={3},
  year={2018},
  url={https://s3-us-west-2.amazonaws.com/openai-assets/research-covers/language-unsupervised/language_understanding_paper.pdf}
}

@article{diffusionModel,
  title={Denoising diffusion probabilistic models},
  author={Ho, Jonathan and Jain, Ajay and Abbeel, Pieter},
  journal={Advances in neural information processing systems},
  volume={33},
  pages={6840--6851},
  year={2020}
}

@article{zhangLFDP,
  title = {Deep Potential Molecular Dynamics: A Scalable Model with the Accuracy of Quantum Mechanics},
  author = {Zhang, Linfeng and Han, Jiequn and Wang, Han and Car, Roberto and E, Weinan},
  journal = {Phys. Rev. Lett.},
  volume = {120},
  issue = {14},
  pages = {143001},
  numpages = {6},
  year = {2018},
  month = {Apr},
  publisher = {American Physical Society},
  doi = {10.1103/PhysRevLett.120.143001},
  url = {https://link.aps.org/doi/10.1103/PhysRevLett.120.143001}
}

@article{KLD,
  title={On information and sufficiency},
  author={Kullback, Solomon and Leibler, Richard A},
  journal={The annals of mathematical statistics},
  volume={22},
  number={1},
  pages={79--86},
  year={1951},
  publisher={JSTOR}
}

@book{deeplearningBook,
    title={Deep Learning},
    author={Ian Goodfellow and Yoshua Bengio and Aaron Courville},
    publisher={MIT Press},
    note={\url{http://www.deeplearningbook.org}},
    year={2016}
}

@article{wuDianPRL,
  title = {Solving Statistical Mechanics Using Variational Autoregressive Networks},
  author = {Wu, Dian and Wang, Lei and Zhang, Pan},
  journal = {Phys. Rev. Lett.},
  volume = {122},
  issue = {8},
  pages = {080602},
  numpages = {6},
  year = {2019},
  month = {Feb},
  publisher = {American Physical Society},
  doi = {10.1103/PhysRevLett.122.080602},
  url = {https://link.aps.org/doi/10.1103/PhysRevLett.122.080602}
}

@article{normalizingflow1,
  title={Normalizing flows for probabilistic modeling and inference},
  author={Papamakarios, George and Nalisnick, Eric and Rezende, Danilo Jimenez and Mohamed, Shakir and Lakshminarayanan, Balaji},
  journal={The Journal of Machine Learning Research},
  volume={22},
  number={1},
  pages={2617--2680},
  year={2021},
  publisher={JMLRORG}
}

@inproceedings{normalizingflow2,
  title={Variational inference with normalizing flows},
  author={Rezende, Danilo and Mohamed, Shakir},
  booktitle={International conference on machine learning},
  pages={1530--1538},
  year={2015},
  organization={PMLR}
}

@InProceedings{pixelRCNN,
  title = 	 {Pixel Recurrent Neural Networks},
  author = 	 {van den Oord, Aäron and Kalchbrenner, Nal and Kavukcuoglu, Koray},
  booktitle = 	 {Proceedings of The 33rd International Conference on Machine Learning},
  pages = 	 {1747--1756},
  year = 	 {2016},
  editor = 	 {Balcan, Maria Florina and Weinberger, Kilian Q.},
  volume = 	 {48},
  series = 	 {Proceedings of Machine Learning Research},
  address = 	 {New York, New York, USA},
  month = 	 {20--22 Jun},
  publisher =    {PMLR},
  url = 	 {https://proceedings.mlr.press/v48/oord16.html}
}

@misc{realnvp,
      title={Density estimation using Real NVP}, 
      author={Laurent Dinh and Jascha Sohl-Dickstein and Samy Bengio},
      year={2017},
      eprint={1605.08803},
      archivePrefix={arXiv},
      primaryClass={cs.LG},
      url={https://arxiv.org/abs/1605.08803}, 
}

@article{LWnewPRL,
  title={Deep Variational Free Energy Approach to Dense Hydrogen},
  author={Xie, Hao and Li, Zi-Hang and Wang, Han and Zhang, Linfeng and Wang, Lei},
  journal={Physical Review Letters},
  volume={131},
  number={12},
  pages={126501},
  year={2023},
  publisher={APS}
}

@article{frankNoe,
author={No{\'e}, Frank and Olsson, Simon and K{\"o}hler, Jonas and Wu, Hao},
title = {Boltzmann generators: Sampling equilibrium states of many-body systems with deep learning},
journal = {Science},
volume = {365},
number = {6457},
pages = {eaaw1147},
year = {2019},
doi = {10.1126/science.aaw1147}
}

@article{atomicNF1,
  title={Normalizing flows for atomic solids},
  author={Wirnsberger, Peter and Papamakarios, George and Ibarz, Borja and Racani{\`e}re, S{\'e}bastien and Ballard, Andrew J and Pritzel, Alexander and Blundell, Charles},
  journal={Machine Learning: Science and Technology},
  volume={3},
  number={2},
  pages={025009},
  year={2022},
  publisher={IOP Publishing}
}

@article{atomicNF2,
  title={Free energy calculation of crystalline solids using normalizing flows},
  author={Ahmad, Rasool and Cai, Wei},
  journal={Modelling and Simulation in Materials Science and Engineering},
  volume={30},
  number={6},
  pages={065007},
  year={2022},
  publisher={IOP Publishing}
}

@article{cubic,
  title={Cubic-spline flows},
  author={Durkan, Conor and Bekasov, Artur and Murray, Iain and Papamakarios, George},
  journal={arXiv preprint arXiv:1906.02145},
  year={2019}
}

@article{fritsch1980monotone,
  title={Monotone piecewise cubic interpolation},
  author={Fritsch, Frederick N and Carlson, Ralph E},
  journal={SIAM Journal on Numerical Analysis},
  volume={17},
  number={2},
  pages={238--246},
  year={1980},
  publisher={SIAM}
}

@article{steffen1990simple,
  title={A simple method for monotonic interpolation in one dimension},
  author={Steffen, Matthias},
  journal={Astronomy and Astrophysics, Vol. 239, NO. NOV (II), P. 443, 1990},
  volume={239},
  pages={443},
  year={1990}
}

@article{equivariantFlowNoe,
  title={Equivariant flows: sampling configurations for multi-body systems with symmetric energies},
  author={K{\"o}hler, Jonas and Klein, Leon and No{\'e}, Frank},
  journal={arXiv preprint arXiv:1910.00753},
  year={2019}
}

@article{equivariantFlowLattice,
  title={Equivariant flow-based sampling for lattice gauge theory},
  author={Kanwar, Gurtej and Albergo, Michael S and Boyda, Denis and Cranmer, Kyle and Hackett, Daniel C and Racaniere, S{\'e}bastien and Rezende, Danilo Jimenez and Shanahan, Phiala E},
  journal={Physical Review Letters},
  volume={125},
  number={12},
  pages={121601},
  year={2020},
  publisher={APS}
}

@article{equivariant1,
  title={Exchangeable neural ode for set modeling},
  author={Li, Yang and Yi, Haidong and Bender, Christopher and Shan, Siyuan and Oliva, Junier B},
  journal={Advances in Neural Information Processing Systems},
  volume={33},
  pages={6936--6946},
  year={2020}
}

@InProceedings{equivariant2,
  title = 	 {Scalable Normalizing Flows for Permutation Invariant Densities},
  author =       {Bilo{\v{s}}, Marin and G{\"u}nnemann, Stephan},
  booktitle = 	 {Proceedings of the 38th International Conference on Machine Learning},
  pages = 	 {957--967},
  year = 	 {2021},
  editor = 	 {Meila, Marina and Zhang, Tong},
  volume = 	 {139},
  series = 	 {Proceedings of Machine Learning Research},
  month = 	 {18--24 Jul},
  publisher =    {PMLR},
  url = 	 {https://proceedings.mlr.press/v139/bilos21a.html}
}

@misc{neuralode,
      title={Neural Ordinary Differential Equations}, 
      author={Ricky T. Q. Chen and Yulia Rubanova and Jesse Bettencourt and David Duvenaud},
      year={2019},
      eprint={1806.07366},
      archivePrefix={arXiv},
      primaryClass={cs.LG},
      url={https://arxiv.org/abs/1806.07366}, 
}

@article{meanfield,
	author = {Jordan, Michael I. and Ghahramani, Zoubin and Jaakkola, Tommi S. and Saul, Lawrence K.},
	date = {1999/11/01},
	doi = {10.1023/A:1007665907178},
	id = {Jordan1999},
	isbn = {1573-0565},
	journal = {Machine Learning},
	number = {2},
	pages = {183--233},
	title = {An Introduction to Variational Methods for Graphical Models},
	url = {https://doi.org/10.1023/A:1007665907178},
	volume = {37},
	year = {1999}}

@misc{vae,
      title={Auto-Encoding Variational Bayes}, 
      author={Diederik P Kingma and Max Welling},
      year={2022},
      eprint={1312.6114},
      archivePrefix={arXiv},
      primaryClass={stat.ML},
      url={https://arxiv.org/abs/1312.6114}, 
}

@misc{gan,
      title={Generative Adversarial Networks}, 
      author={Ian J. Goodfellow and Jean Pouget-Abadie and Mehdi Mirza and Bing Xu and David Warde-Farley and Sherjil Ozair and Aaron Courville and Yoshua Bengio},
      year={2014},
      eprint={1406.2661},
      archivePrefix={arXiv},
      primaryClass={stat.ML},
      url={https://arxiv.org/abs/1406.2661}, 
}

@misc{ffjord,
      title={FFJORD: Free-form Continuous Dynamics for Scalable Reversible Generative Models}, 
      author={Will Grathwohl and Ricky T. Q. Chen and Jesse Bettencourt and Ilya Sutskever and David Duvenaud},
      year={2018},
      eprint={1810.01367},
      archivePrefix={arXiv},
      primaryClass={cs.LG},
      url={https://arxiv.org/abs/1810.01367}, 
}

@InProceedings{inverResNet,
  title = 	 {Invertible Residual Networks},
  author =       {Behrmann, Jens and Grathwohl, Will and Chen, Ricky T. Q. and Duvenaud, David and Jacobsen, Joern-Henrik},
  booktitle = 	 {Proceedings of the 36th International Conference on Machine Learning},
  pages = 	 {573--582},
  year = 	 {2019},
  editor = 	 {Chaudhuri, Kamalika and Salakhutdinov, Ruslan},
  volume = 	 {97},
  series = 	 {Proceedings of Machine Learning Research},
  month = 	 {09--15 Jun},
  publisher =    {PMLR},
  url = 	 {https://proceedings.mlr.press/v97/behrmann19a.html}
}

@inbook{neb1,
author = {HANNES JÓNSSON and GREG MILLS and KARSTEN W. JACOBSEN},
title = {Nudged elastic band method for finding minimum energy paths of transitions},
booktitle = {Classical and Quantum Dynamics in Condensed Phase Simulations},
chapter = {},
pages = {385-404},
doi = {10.1142/9789812839664_0016},
URL = {https://www.worldscientific.com/doi/abs/10.1142/9789812839664_0016}
}

@article{neb2,
title = {Reversible work transition state theory: application to dissociative adsorption of hydrogen},
journal = {Surface Science},
volume = {324},
number = {2},
pages = {305-337},
year = {1995},
issn = {0039-6028},
doi = {https://doi.org/10.1016/0039-6028(94)00731-4},
url = {https://www.sciencedirect.com/science/article/pii/0039602894007314},
author = {Gregory Mills and Hannes Jónsson and Gregory K. Schenter}
}

@inproceedings{glow,
 author = {Kingma, Durk P and Dhariwal, Prafulla},
 booktitle = {Advances in Neural Information Processing Systems},
 editor = {S. Bengio and H. Wallach and H. Larochelle and K. Grauman and N. Cesa-Bianchi and R. Garnett},
 pages = {},
 publisher = {Curran Associates, Inc.},
 title = {Glow: Generative Flow with Invertible 1x1 Convolutions},
 url = {https://proceedings.neurips.cc/paper_files/paper/2018/file/d139db6a236200b21cc7f752979132d0-Paper.pdf},
 volume = {31},
 year = {2018}
}

@article{tica1,
  title = {Separation of a mixture of independent signals using time delayed correlations},
  author = {Molgedey, L. and Schuster, H. G.},
  journal = {Phys. Rev. Lett.},
  volume = {72},
  issue = {23},
  pages = {3634--3637},
  numpages = {0},
  year = {1994},
  month = {Jun},
  publisher = {American Physical Society},
  doi = {10.1103/PhysRevLett.72.3634},
  url = {https://link.aps.org/doi/10.1103/PhysRevLett.72.3634}
}

@article{tica2,
    author = {Pérez-Hernández, Guillermo and Paul, Fabian and Giorgino, Toni and De Fabritiis, Gianni and Noé, Frank},
    title = {Identification of slow molecular order parameters for Markov model construction},
    journal = {The Journal of Chemical Physics},
    volume = {139},
    number = {1},
    pages = {015102},
    year = {2013},
    month = {07},
    issn = {0021-9606},
    doi = {10.1063/1.4811489},
    url = {https://doi.org/10.1063/1.4811489},
}

@article{atb1,
author = {Malde, Alpeshkumar K. and Zuo, Le and Breeze, Matthew and Stroet, Martin and Poger, David and Nair, Pramod C. and Oostenbrink, Chris and Mark, Alan E.},
title = {An Automated Force Field Topology Builder (ATB) and Repository: Version 1.0},
journal = {Journal of Chemical Theory and Computation},
volume = {7},
number = {12},
pages = {4026-4037},
year = {2011},
doi = {10.1021/ct200196m},
note ={PMID: 26598349},
URL = {https://doi.org/10.1021/ct200196m},
eprint = {https://doi.org/10.1021/ct200196m}
}

@article{atb3,
author = {Stroet, Martin and Caron, Bertrand and Visscher, Koen M. and Geerke, Daan P. and Malde, Alpeshkumar K. and Mark, Alan E.},
title = {Automated Topology Builder Version 3.0: Prediction of Solvation Free Enthalpies in Water and Hexane},
journal = {Journal of Chemical Theory and Computation},
volume = {14},
number = {11},
pages = {5834-5845},
year = {2018},
doi = {10.1021/acs.jctc.8b00768},
note ={PMID: 30289710},
URL = {https://doi.org/10.1021/acs.jctc.8b00768},
eprint = {https://doi.org/10.1021/acs.jctc.8b00768}
}

@article{54A7,
author={Schmid, Nathan
and Eichenberger, Andreas P.
and Choutko, Alexandra
and Riniker, Sereina
and Winger, Moritz
and Mark, Alan E.
and van Gunsteren, Wilfred F.},
title={Definition and testing of the GROMOS force-field versions 54A7 and 54B7},
journal={European Biophysics Journal},
year={2011},
month={Jul},
day={01},
volume={40},
number={7},
pages={843-856},
issn={1432-1017},
doi={10.1007/s00249-011-0700-9},
url={https://doi.org/10.1007/s00249-011-0700-9}
}

@article{chignolin,
  title={Folding free-energy landscape of a 10-residue mini-protein, chignolin},
  author={Satoh, Daisuke and Shimizu, Kentaro and Nakamura, Shugo and Terada, Tohru},
  journal={FEBS letters},
  volume={580},
  number={14},
  pages={3422--3426},
  year={2006},
  publisher={Wiley Online Library}
}

@article{1uao,
  title={Mutation-induced change in chignolin stability from $\pi$-turn to $\alpha$-turn},
  author={Maruyama, Yutaka and Koroku, Shunpei and Imai, Misaki and Takeuchi, Koh and Mitsutake, Ayori},
  journal={RSC advances},
  volume={10},
  number={38},
  pages={22797--22808},
  year={2020},
  publisher={Royal Society of Chemistry}
}

@article{1uao2,
	author = {Fischer, Anna-Lena M. and Tichy, Anna and Kokot, Janik and Hoerschinger, Valentin J. and Wild, Robert F. and Riccabona, Jakob R. and Loeffler, Johannes R. and Waibl, Franz and Quoika, Patrick K. and Gschwandtner, Philipp and Forli, Stefano and Ward, Andrew B. and Liedl, Klaus R. and Zacharias, Martin and Fern{\'a}ndez-Quintero, Monica L.},
	date = {2024/03/12},
	doi = {10.1021/acs.jctc.3c01106},
	isbn = {1549-9618},
	journal = {Journal of Chemical Theory and Computation},
	journal1 = {Journal of Chemical Theory and Computation},
	journal2 = {J. Chem. Theory Comput.},
	month = {03},
	number = {5},
	pages = {2321--2333},
	publisher = {American Chemical Society},
	title = {The Role of Force Fields and Water Models in Protein Folding and Unfolding Dynamics},
	type = {doi: 10.1021/acs.jctc.3c01106},
	url = {https://doi.org/10.1021/acs.jctc.3c01106},
	volume = {20},
	year = {2024},
	year1 = {2024}}

@inproceedings{resnet,
  title={Deep residual learning for image recognition},
  author={He, Kaiming and Zhang, Xiangyu and Ren, Shaoqing and Sun, Jian},
  booktitle={Proceedings of the IEEE conference on computer vision and pattern recognition},
  pages={770--778},
  year={2016}
}

@article{addH,
	author = {Kunzmann, Patrick and Anter, Jacob Marcel and Hamacher, Kay},
	date = {2022/03/29},
	doi = {10.1186/s13015-022-00215-x},
	id = {Kunzmann2022},
	isbn = {1748-7188},
	journal = {Algorithms for Molecular Biology},
	number = {1},
	pages = {7},
	title = {Adding hydrogen atoms to molecular models via fragment superimposition},
	url = {https://doi.org/10.1186/s13015-022-00215-x},
	volume = {17},
	year = {2022},
    }

@article{CA3CA8,
	author = {Kang, Peilin and Trizio, Enrico and Parrinello, Michele},
	date = {2024/06/01},
	doi = {10.1038/s43588-024-00645-0},
	id = {Kang2024},
	isbn = {2662-8457},
	journal = {Nature Computational Science},
	number = {6},
	pages = {451--460},
	title = {Computing the committor with the committor to study the transition state ensemble},
	url = {https://doi.org/10.1038/s43588-024-00645-0},
	volume = {4},
	year = {2024}}

@article{folding1,
  title={Folding dynamics of 10-residue $\beta$-hairpin peptide chignolin},
  author={Suenaga, Atsushi and Narumi, Tetsu and Futatsugi, Noriyuki and Yanai, Ryoko and Ohno, Yousuke and Okimoto, Noriaki and Taiji, Makoto},
  journal={Chemistry--An Asian Journal},
  volume={2},
  number={5},
  pages={591--598},
  year={2007},
  publisher={Wiley Online Library}
}

@article{folding2,
  title={Exploring the folding free energy landscape of a $\beta$-hairpin miniprotein, chignolin, using multiscale free energy landscape calculation method},
  author={Harada, Ryuhei and Kitao, Akio},
  journal={The Journal of Physical Chemistry B},
  volume={115},
  number={27},
  pages={8806--8812},
  year={2011},
  publisher={ACS Publications}
}

@article{folding3,
  title={Understanding ensemble protein folding at atomic detail},
  author={Hubner, Isaac A and Deeds, Eric J and Shakhnovich, Eugene I},
  journal={Proceedings of the National Academy of Sciences},
  volume={103},
  number={47},
  pages={17747--17752},
  year={2006},
  publisher={National Academy of Sciences}
}

@article{folding4,
  title={How fast-folding proteins fold},
  author={Lindorff-Larsen, Kresten and Piana, Stefano and Dror, Ron O and Shaw, David E},
  journal={Science},
  volume={334},
  number={6055},
  pages={517--520},
  year={2011},
  publisher={American Association for the Advancement of Science}
}

@article{conditionalBG,
  title={Efficient mapping of phase diagrams with conditional {Boltzmann} generators},
  author={Schebek, Maximilian and Invernizzi, Michele and No{\'e}, Frank and Rogal, Jutta},
  journal={Machine Learning: Science and Technology},
  volume={5},
  number={4},
  pages={045045},
  year={2024},
  doi={10.1088/2632-2153/ad849d}
}

@article{conditionedBGtps,
  title={Conditioning {Boltzmann} generators for rare event sampling},
  author={Falkner, Sebastian and Coretti, Alessandro and Romano, Salvatore and Geissler, Phillip L and Dellago, Christoph},
  journal={Machine Learning: Science and Technology},
  volume={4},
  number={3},
  pages={035050},
  year={2023},
  doi={10.1088/2632-2153/acf55c}
}

@article{DiG,
  title={Predicting equilibrium distributions for molecular systems with deep learning},
  author={Zheng, Shuxin and He, Jiyan and Liu, Chang and Shi, Yu and Lu, Ziheng and Feng, Weitao and Ju, Fusong and Wang, Jiaxi and Zhu, Jianwei and Min, Yaosen and Zhang, He and Tang, Shidi and Hao, Hongxia and Jin, Peiran and Chen, Chi and No{\'e}, Frank and Liu, Haiguang and Liu, Tie-Yan},
  journal={Nature Machine Intelligence},
  volume={6},
  pages={558--567},
  year={2024},
  doi={10.1038/s42256-024-00837-3}
}

@article{BioEmu,
  title={Scalable emulation of protein equilibrium ensembles with generative deep learning},
  author={Lewis, Sarah and Hempel, Tim and Jimenez-Luna, Jose and Gastegger, Michael and Xie, Yu and Foong, Andrew Y K and Garcia Satorras, Victor and Abdin, Osama and Veeling, Bastiaan S and Zaporozhets, Iryna and Chen, Yaoyi and Yang, Soojung and Schneuing, Arne and Nigam, Jigyasa and Barbero, Federico and Stimper, Vincent and Campbell, Andrew and Yim, Jason and Lienen, Marten and Shi, Yu and Zheng, Shuxin and Schulz, Hannes and Munir, Usman and Clementi, Cecilia and No{\'e}, Frank},
  journal={Science},
  volume={387},
  pages={eadv9817},
  year={2025},
  doi={10.1126/science.adv9817}
}

@article{enhancedDiffusion,
  title={Enhanced diffusion sampling: efficient rare event sampling and free energy calculation with diffusion models},
  author={Xie, Yu and Winkler, Ludwig and Sun, Lixin and Lewis, Sarah and Foster, Adam E and Jimenez-Luna, Jos{\'e} and Hempel, Tim and Gastegger, Michael and Chen, Yaoyi and Zaporozhets, Iryna and Clementi, Cecilia and Bishop, Christopher M and No{\'e}, Frank},
  journal={arXiv preprint arXiv:2602.16634},
  year={2026}
}

@article{alphaFold3,
	author = {Abramson, Josh and Adler, Jonas and Dunger, Jack and Evans, Richard and Green, Tim and Pritzel, Alexander and Ronneberger, Olaf and Willmore, Lindsay and Ballard, Andrew J. and Bambrick, Joshua and Bodenstein, Sebastian W. and Evans, David A. and Hung, Chia-Chun and O'Neill, Michael and Reiman, David and Tunyasuvunakool, Kathryn and Wu, Zachary and {\v Z}emgulyt{\.e}, Akvil{\.e} and Arvaniti, Eirini and Beattie, Charles and Bertolli, Ottavia and Bridgland, Alex and Cherepanov, Alexey and Congreve, Miles and Cowen-Rivers, Alexander I. and Cowie, Andrew and Figurnov, Michael and Fuchs, Fabian B. and Gladman, Hannah and Jain, Rishub and Khan, Yousuf A. and Low, Caroline M. R. and Perlin, Kuba and Potapenko, Anna and Savy, Pascal and Singh, Sukhdeep and Stecula, Adrian and Thillaisundaram, Ashok and Tong, Catherine and Yakneen, Sergei and Zhong, Ellen D. and Zielinski, Michal and {\v Z}{\'\i}dek, Augustin and Bapst, Victor and Kohli, Pushmeet and Jaderberg, Max and Hassabis, Demis and Jumper, John M.},
	date = {2024/06/01},
	doi = {10.1038/s41586-024-07487-w},
	id = {Abramson2024},
	isbn = {1476-4687},
	journal = {Nature},
	number = {8016},
	pages = {493--500},
	title = {Accurate structure prediction of biomolecular interactions with AlphaFold 3},
	url = {https://doi.org/10.1038/s41586-024-07487-w},
	volume = {630},
	year = {2024}}

@misc{simpleFold,
      title={SimpleFold: Folding Proteins is Simpler than You Think}, 
      author={Yuyang Wang and Jiarui Lu and Navdeep Jaitly and Josh Susskind and Miguel Angel Bautista},
      year={2025},
      eprint={2509.18480},
      archivePrefix={arXiv},
      primaryClass={cs.LG},
      url={https://arxiv.org/abs/2509.18480}, 
}

@inproceedings{equivariantNet,
  title={E (n) equivariant graph neural networks},
  author={Satorras, V{\i}ctor Garcia and Hoogeboom, Emiel and Welling, Max},
  booktitle={International conference on machine learning},
  pages={9323--9332},
  year={2021},
  organization={PMLR}
}

@article{klein2023equivariant,
  title={Equivariant flow matching},
  author={Klein, Leon and Kr{\"a}mer, Andreas and No{\'e}, Frank},
  journal={Advances in Neural Information Processing Systems},
  volume={36},
  pages={59886--59910},
  year={2023}
}

@article{winter2022unsupervised,
  title={Unsupervised learning of group invariant and equivariant representations},
  author={Winter, Robin and Bertolini, Marco and Le, Tuan and Noe, Frank and Clevert, Djork-Arn{\'e}},
  journal={Advances in Neural Information Processing Systems},
  volume={35},
  pages={31942--31956},
  year={2022}
}

@article{genAI4chem,
author = {Pratyush Tiwary  and Lukas Herron  and Richard John  and Suemin Lee  and Disha Sanwal  and Ruiyu Wang },
title = {Generative AI for computational chemistry: A roadmap to predicting emergent phenomena},
journal = {Proceedings of the National Academy of Sciences},
volume = {122},
number = {41},
pages = {e2415655121},
year = {2025},
doi = {10.1073/pnas.2415655121},
URL = {https://www.pnas.org/doi/abs/10.1073/pnas.2415655121},
eprint = {https://www.pnas.org/doi/pdf/10.1073/pnas.2415655121}}

@misc{genAI4chem2,
      title={Combined Representation and Generation with Diffusive State Predictive Information Bottleneck}, 
      author={Richard John and Yunrui Qiu and Lukas Herron and Pratyush Tiwary},
      year={2025},
      eprint={2510.09784},
      archivePrefix={arXiv},
      primaryClass={cs.LG},
      url={https://arxiv.org/abs/2510.09784}, 
}

@article{genAI4chem3,
  title = {Inferring the Isotropic-Nematic Phase Transition with Generative Machine Learning},
  author = {Beyerle, Eric R. and Tiwary, Pratyush},
  journal = {Phys. Rev. Lett.},
  volume = {135},
  issue = {6},
  pages = {068102},
  numpages = {7},
  year = {2025},
  month = {Aug},
  publisher = {American Physical Society},
  doi = {10.1103/1wdj-ym3s},
  url = {https://link.aps.org/doi/10.1103/1wdj-ym3s}
}

@article{chemDL,
	author = {Cersonsky, Rose K. and Cheng, Bingqing and De Vivo, Marco and Tiwary, Pratyush},
	date = {2025/06/10},
	doi = {10.1021/acs.jctc.5c00650},
	isbn = {1549-9618},
	journal = {Journal of Chemical Theory and Computation},
	journal1 = {Journal of Chemical Theory and Computation},
	journal2 = {J. Chem. Theory Comput.},
	month = {06},
	number = {11},
	pages = {5359--5364},
	publisher = {American Chemical Society},
	title = {Machine Learning and Statistical Mechanics: Shared Synergies for Next Generation of Chemical Theory and Computation},
	type = {doi: 10.1021/acs.jctc.5c00650},
	url = {https://doi.org/10.1021/acs.jctc.5c00650},
	volume = {21},
	year = {2025},
	year1 = {2025}}

@article{chemDL2,
	author = {Goonetilleke, Eshani C. and Liu, Bojun and Wu, Yue and O'Connor, Michael S. and Huang, Xuhui},
	date = {2025/11/27},
	doi = {10.1021/acs.jpcb.5c06097},
	isbn = {1520-6106},
	journal = {The Journal of Physical Chemistry B},
	journal1 = {The Journal of Physical Chemistry B},
	journal2 = {J. Phys. Chem. B},
	month = {11},
	number = {47},
	pages = {12133--12145},
	publisher = {American Chemical Society},
	title = {A Practical Guide to Transition State Analysis in Biomolecular Simulations with TS-DAR},
	type = {doi: 10.1021/acs.jpcb.5c06097},
	url = {https://doi.org/10.1021/acs.jpcb.5c06097},
	volume = {129},
	year = {2025},
	year1 = {2025}}

@article{aicv1,
	author = {Liu, Bojun and Boysen, Jordan G. and Unarta, Ilona Christy and Du, Xuefeng and Li, Yixuan and Huang, Xuhui},
	date = {2025/01/02},
	doi = {10.1038/s41467-024-55228-4},
	id = {Liu2025},
	isbn = {2041-1723},
	journal = {Nature Communications},
	number = {1},
	pages = {349},
	title = {Exploring transition states of protein conformational changes via out-of-distribution detection in the hyperspherical latent space},
	url = {https://doi.org/10.1038/s41467-024-55228-4},
	volume = {16},
	year = {2025}}

@article{aicv2,
	author = {Liu, Bojun and Cao, Siqin and Boysen, Jordan G. and Xue, Mingyi and Huang, Xuhui},
	date = {2025/07/01},
	doi = {10.1038/s43588-025-00815-8},
	id = {Liu2025},
	isbn = {2662-8457},
	journal = {Nature Computational Science},
	number = {7},
	pages = {562--571},
	title = {Memory kernel minimization-based neural networks for discovering slow collective variables of biomolecular dynamics},
	url = {https://doi.org/10.1038/s43588-025-00815-8},
	volume = {5},
	year = {2025}}

@article{aicv3,
  title={GraphVAMPnets for uncovering slow collective variables of self-assembly dynamics},
  author={Liu, Bojun and Xue, Mingyi and Qiu, Yunrui and Konovalov, Kirill A and O’Connor, Michael S and Huang, Xuhui},
  journal={The Journal of Chemical Physics},
  volume={159},
  number={9},
  year={2023},
  publisher={AIP Publishing}
}

@article{stringMethod,
  title = {String method for the study of rare events},
  author = {E, Weinan and Ren, Weiqing and Vanden-Eijnden, Eric},
  journal = {Phys. Rev. B},
  volume = {66},
  issue = {5},
  pages = {052301},
  numpages = {4},
  year = {2002},
  month = {Aug},
  publisher = {American Physical Society},
  doi = {10.1103/PhysRevB.66.052301},
  url = {https://link.aps.org/doi/10.1103/PhysRevB.66.052301}
}
\appendix

\clearpage
\pagebreak
\widetext
\begin{center}
  \textbf{Supplementary Information}
\end{center}

\setcounter{equation}{0}
\setcounter{figure}{0}
\setcounter{table}{0}
\setcounter{algocf}{0}

\renewcommand{\theequation}{S\arabic{equation}}
\renewcommand{\thefigure}{S\arabic{figure}}
\renewcommand{\thetable}{S\arabic{table}}
\renewcommand{\thealgocf}{S\arabic{algocf}}

\makeatletter
\@removefromreset{equation}{section}
\@removefromreset{figure}{section}
\@removefromreset{table}{section}
\makeatother

\section{Supplementary for the bistable dimer potential application}

\subsection{Technical details in training}

The bistable dimer potential of Eq.~(\ref{eq:dimerH}) acts on a $3$-dimensional vector $\bm{x}=(x_0,\,x_1,\,x_2)$ connecting the two particles.
Its bijective transformation maps $\bm{x}$ to $(|\bm{x}|,\, u_0,\, u_1)$.
$|\bm{x}|$ is the bond-length CV in a range of $[1,\, 6]$ and the two auxiliary variables $u_0,\,u_1$ are sampled from a uniform prior on $[0,\, 1]$.
In a second training, temperature $T$ was included as an additional environmental parameter, uniformly sampled from $[0.3,\, 1.6]$, so that one optimization yields a temperature-continuous FES estimate.
The cubic-spline flow model used in these two VaFES optimizations is the same, and has the same structure as the one introduced in \cite{cubic} with the parameter neural network being an MLP network.
A typical checkerboard partition is used for the coupling layers of the cubic-spline flow.
The remaining hyperparameters are listed in Table~\ref{tab:dimer}.

\begin{table}[h]
\centering
\caption{VaFES hyperparameters for the bistable dimer potential application.}
\label{tab:dimer}
\begin{tabular}{l l}
\hline\hline
Parameter & Value \\
\hline
Spline bins $K$ & 50 \\
Coupling layers & 16 \\
  MLP hidden dimensions & $20,\, 50,\, 100,\, 150$ \\
  Activation & ELU\\
Prior distribution & Uniform \\
Optimizer & Adamax \\
Learning rate & $7\times 10^{-4}$ \\
LR schedule & StepLR (step $1000$, $\gamma=0.8$) \\
Batch size & 512 \\
Bond-length CV range & $[1,\, 6]$ \\
Temperature $T$ range & $[0.3,\, 1.6]$ \\
\hline\hline
\end{tabular}
\end{table}

\subsection{Details about the reversible transformation and its Jacobian}
For the bistable dimer potential, the reversible CV transformation used is as follows
\begin{equation}
  \begin{aligned}
    (x_0,\, x_1,\, x_2) &\xmapsto{\;\mathcal{T}\;} (|\bm{x}|,\, u_0,\, u_1), \\
    \text{with}\ |\bm{x}| &= \sqrt{x_0^2 + x_1^2 + x_2^2}, \\
    u_0 &= \sign(x_1)\frac{x_1^2}{x_0^2+x_1^2}, \\
    u_1 &= \sign(x_2)\frac{x_2^2}{x_0^2+x_1^2+x_2^2}.
  \end{aligned}
\end{equation}
Notably, this transformation is not bijective, but it is reversible within the variable ranges used in training.
A proper bijective distance computation admits a more complex form, and is therefore postponed to the chignolin application.

The corresponding Jacobian matrix, $\partial(|\bm{x}|,u_0,u_1)/\partial(x_0,x_1,x_2)$, is
\begin{equation}
  \bm{J}=
  \begin{pmatrix}
    \dfrac{x_0}{|\bm{x}|} & \dfrac{x_1}{|\bm{x}|} & \dfrac{x_2}{|\bm{x}|} \\
    -2\sign(x_1)\dfrac{x_0 x_1^2}{(x_0^2+x_1^2)^2} & 2\sign(x_1)\dfrac{x_0^2 x_1}{(x_0^2+x_1^2)^2} & 0 \\
    -2\sign(x_2)\dfrac{x_0 x_2^2}{|\bm{x}|^4} & -2\sign(x_2)\dfrac{x_1 x_2^2}{|\bm{x}|^4} & 2\sign(x_2)\dfrac{x_2(x_0^2+x_1^2)}{|\bm{x}|^4}
  \end{pmatrix}.
  \label{eq:dimerJac}
\end{equation}

\subsection{Analytical results of the FES as a continuous function of the temperature}
\begin{figure*}[!htbp]
\centering
\includegraphics[width=1\textwidth]{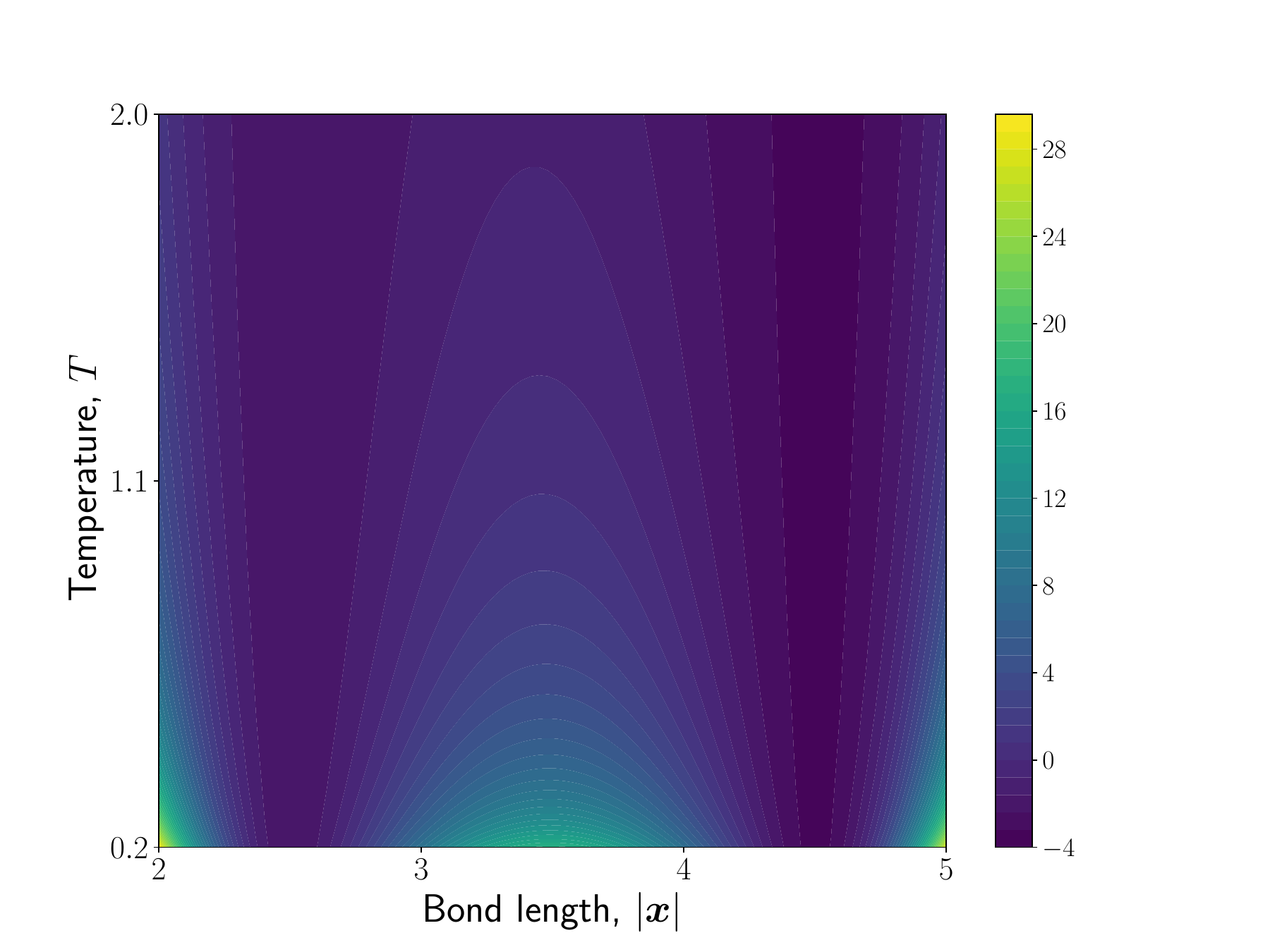}
\caption{
  The analytical results of the FES of the bistable dimer potential system as a continuous function of the temperature and the bond length CV.
}
\label{fig:dimerExact}
\end{figure*}

As a ground truth for comparison, we show the analytical results of the FES of the bistable dimer potential as a continuous function of the temperature in Fig.~\ref{fig:dimerExact}.

\section{Supplementary for the diazene application}

\subsection{Technical details in training}

We used the Gromos 54A7 force field~\cite{54A7}. The force field parameters of \ce{N2H2} were generated using the Automated Topology Builder (ATB)~\cite{atb1, atb3}. The thermodynamic temperature is set to approximately $120$ K.
Throughout the atomistic applications, energies are measured in kJ/mol, so the temperature variable $T$ in VaFES denotes the thermal energy $k_B T_{\mathrm{phys}}$ expressed in kJ/mol rather than the temperature in kelvin. For $120$ K, this gives $T \approx 1.0\si{\kilo\joule\per\mol}$.

The machine-learning CV $s_1$ was trained by a single-layer sigmoid-based transformation with an MLP of hidden dimensions $[10,\, 10]$.
The training used MD trajectory data of cis and trans configurations, split into training and test sets at a ratio of $0.8$, and optimized for $100$ epochs with a batch size of $256$.

The model architecture of the cubic-spline flow model used in VaFES is the same as in \cite{cubic} with the parameter neural network being an MLP network.
A typical checkerboard partition is used for the coupling layers of the cubic-spline flow.
The ranges of each dimension were determined from the data collected in the machine-learning CV training.
The hyperparameters are listed in Table~\ref{tab:h2n2}.

\begin{table}[h]
\centering
\caption{Hyperparameters for the diazene application.}
\label{tab:h2n2}
\begin{tabular}{l l}
\hline\hline
Parameter & Value \\
\hline
Spline bins $K$ & 50 \\
Coupling layers & 16 \\
MLP hidden dimensions & $(20,\, 50,\, 100,\, 150)$ \\
Activation & ELU\\
Prior distribution & Uniform \\
Optimizer & Adamax \\
Learning rate & $7\times 10^{-4}$ \\
LR schedule & StepLR (step $1000$, $\gamma=0.8$) \\
Batch size & 512 \\
\hline\hline
\end{tabular}
\end{table}

\subsection{Details about the bijective transformation and its Jacobian}

For the diazene application, from the $6$-dimensional state, $(x_1,\, y_1,\, d,\, x_2,\, y_2,\, z_2)$, we applied an additional bijective transformation on $y_2$ to define the machine-learning CV $s_1\in[0,1]$.
We first rescale $y_2\in[-0.18,0.18]$ to $\tilde{y}_2\in[0,1]$ using a linear transformation, and then apply
\begin{equation}
  \begin{aligned}
    s_1 &= L\,\sigma\!
    \left(k(\tilde{y}_2-t)\right) + b,\\
    \text{with}\ \sigma(x)=\frac{1}{1+e^{-x}}, \qquad
    L &= \frac{(1+e^{k t})(1+e^{k(1-t)})}{e^{k}-1},\qquad
    b = \frac{1+e^{k(1-t)}}{1-e^{k}},
  \end{aligned}
  \label{eq:h2n2Sigmoid}
\end{equation}
where $k$ and $t$ are parameters from the output of an MLP network with $(x_1,\, y_1,\, d,\, x_2,\, z_2)$ as its input.
We then ensure $k>0$ and $t\in[0,1]$, implemented correspondingly by softplus and sigmoid.
This formulation of $L$ and $b$ in Eq.~(\ref{eq:h2n2Sigmoid}) also enforces $s_1(\tilde{y}_2=0)=0$ and $s_1(\tilde{y}_2=1)=1$.

Then, the end-to-end bijective transformation to the intermediate representation latent space is as follows
\begin{equation}
  (x_1,\, y_1,\, d,\, x_2,\, y_2,\, z_2) \xmapsto{\;\mathcal{T}\;} (x_1,\, y_1,\, d,\, x_2,\, s_1,\, z_2).
\end{equation}
Because all variables other than $y_2$ are unchanged by $\mathcal{T}$, the Jacobian matrix is lower triangular with unit diagonal entries except for the $y_2\mapsto s_1$ entry.
Its determinant therefore reduces to
\begin{equation}
  \det \bm{J} = \frac{\partial s_1}{\partial y_2} = \frac{Lk}{0.36}\,\sigma\!\left(k(\tilde{y}_2-t)\right)\left[1-\sigma\!\left(k(\tilde{y}_2-t)\right)\right].
  \label{eq:h2n2Jac}
\end{equation}

\section{Supplementary for the alanine dipeptide application}

\subsection{Technical details in training}

The parameters of the classical force field used for the alanine dipeptide molecule are from the \texttt{Amber99SB} force field. The thermodynamic temperature is set at $300$ K.
As above, energies are measured in kJ/mol, so the temperature variable $T$ in VaFES denotes the thermal energy $k_B T_{\mathrm{phys}}$ in kJ/mol. For $300$ K, this gives $T \approx 2.49\si{\kilo\joule\per\mol}$.

The MD trajectory used for TICA was first converted from the $13$-atom representation into the $33$-dimensional concise representation.
Before applying TICA, we whitened this $33$-dimensional representation using its covariance matrix.
TICA was then performed in the whitened space, so that the end-to-end transformation from the $33$-dimensional coordinates to the TICA coordinates is a full-rank orthogonal transformation.
The first two components were selected as CVs, and the boundary ranges for all $33$ components were determined from the data.
The prior distributions for the $31$ auxiliary dimensions are uniform distributions over the ranges determined from the data.
During training, the two CV dimensions are independently and uniformly sampled over their respective ranges.

The model architecture of the cubic-spline flow model used in VaFES is the same as in \cite{cubic} with the parameter neural network being an MLP network.
A typical checkerboard partition is used for the coupling layers of the cubic-spline flow.
The hyperparameters are listed in Table~\ref{tab:aldp}.

\begin{table}[h]
\centering
\caption{Hyperparameters for the alanine dipeptide application.}
\label{tab:aldp}
\begin{tabular}{l l}
\hline\hline
Parameter & Value \\
\hline
Temperature & $2.49\si{\kilo\joule\per\mol}\ (300\,\mathrm{K})$ \\
Spline bins $K$ & 50 \\
Coupling layers & 12 \\
MLP hidden dimensions & $(512,\, 512,\, 512,\, 1024,\, 1024)$ \\
Activation & ELU\\
Prior distribution & Uniform \\
Optimizer & Adamax \\
Learning rate & $7\times 10^{-4}$ \\
LR schedule & StepLR (step $1000$, $\gamma=0.8$) \\
Batch size & 512 \\
\hline\hline
\end{tabular}
\end{table}

\section{Supplementary for the chignolin application}

\subsection{Details about the bijective transformation and its Jacobian}

The full-atom Cartesian coordinates of the chignolin protein are converted into an intermediate representation by the sequential application of four reversible transformations, each equipped with a closed-form log-Jacobian determinant.
The four transformations are described in the following sections, followed by a description of how these four bijective transformations compose the overall bijective transformation to the latent space of intermediate representation.

\subsubsection{Bijective version of the spherical coordinate transformation}

Given a displacement vector $\bm{x}=(x,y,z)$ between two reference atoms, the modified spherical coordinate transformation maps it to $(r,a,b)$ defined by
\begin{equation}
    r = \sqrt{x^2+y^2+z^2},\qquad
    a = \frac{z}{r},\qquad
    b = \operatorname{atan2}(y,\,x).
\end{equation}
Here $r$ is the inter-atomic distance, $a\in[-1,1]$ is the normalized $z$-projection, \ie, the cosine of the polar angle, and $b\in(-\pi,\pi]$ is the azimuthal angle.
The Jacobian matrix, $\bm{J} = \partial(r,a,b)/\partial(x,y,z)$, can be computed analytically, the log-absolute-determinant of which is
\begin{equation}
  \log\lvert\det \bm{J}\rvert = -2\log r.
\end{equation}

\subsubsection{Bijective version of the local-frame coordinate transformation}

Given a set of $N\ge3$ atoms with positions $\bm{p}_1,\dots,\bm{p}_N$, the first three points $A=\bm{p}_1$, $B=\bm{p}_2$, $C=\bm{p}_3$ define a local orthonormal frame as follows.
Let
\begin{equation}
  \hat{\bm{u}} = \frac{B-A}{\lVert B-A\rVert},\qquad
  \bm{w} = (C-A) - \bigl[(C-A)\cdot\hat{\bm{u}}\bigr]\hat{\bm{u}},\qquad
  \hat{\bm{v}} = \frac{\bm{w}}{\lVert\bm{w}\rVert},\qquad
  \hat{\bm{w}} = \hat{\bm{u}}\times\hat{\bm{v}}.
\end{equation}
The rotation matrix $R=[\hat{\bm{u}},\hat{\bm{v}},\hat{\bm{w}}]$ is parameterized by Euler angles $(\alpha,\beta,\gamma)$ with the ZYX convention.
The frame geometry is captured by three scalars
\begin{equation}
  s_1 = \lVert B-A\rVert,\qquad
  p = (C-A)\cdot\hat{\bm{u}},\qquad
  s = \lVert(C-A) - p\,\hat{\bm{u}}\rVert,
\end{equation}
where $s_1$ is the distance $AB$, $p$ is the projection of $AC$ onto $\hat{\bm{u}}$, and $s$ is the perpendicular distance of $C$ from the line $AB$.
Every remaining atom $\bm{p}_i$ ($i>3$) is expressed in the local frame as $\bm{q}_i = R^\top(\bm{p}_i - A)$.

The complete output of the transformation is
\begin{equation}
  (\bm{p}_1,\dots,\bm{p}_N) \;\xmapsto{\;\mathcal{T}\;}\;
  (A_x,A_y,A_z,\;\alpha,\beta,\gamma,\;s_1,p,s,\;\bm{q}_4,\dots,\bm{q}_N).
\end{equation}
The log-absolute-determinant of the Jacobian is
\begin{equation}
  \log\lvert\det \bm{J}\rvert = -2\log s_1 - \log s - \log\lvert\cos\beta\rvert.
\end{equation}

\subsubsection{Bijective version of the torsion transformation}

Given four ordered atom positions $N,A,B,X$, this transformation replaces the Cartesian position of the terminal atom $X$ with the internal coordinates $(r,\theta,\chi)$ defined by
\begin{equation}
    r = \lVert X-B\rVert,\qquad
    \theta = \arccos\!\left(\frac{(A-B)\cdot(X-B)}{\lVert A-B\rVert\,\lVert X-B\rVert}\right), \qquad
    \chi = \operatorname{atan2}\!\bigl(-\bm{n}_{BA}\cdot(\bm{t}\times\bm{u}),\;\bm{t}\cdot\bm{u}\bigr),
\end{equation}
where $\theta$ is the bond angle $\angle ABX$, $\chi$ is the dihedral (torsion) angle about the $A$--$B$ bond, and
\begin{equation}
  \bm{t} = (A-N)\times(B-A),\qquad \bm{u} = (B-A)\times(X-B),\qquad \bm{n}_{BA} = \frac{B-A}{\lVert B-A\rVert}.
\end{equation}
The positions of $N$, $A$, and $B$ are kept unchanged.
The log-absolute-determinant of the Jacobian is
\begin{equation}
  \log\lvert\det \bm{J}\rvert = -2\log r - \log\sin\theta.
\end{equation}

The inverse transformation reconstructs $X$ from $(r,\theta,\chi)$ by building a local cylindrical frame from $N$, $A$, $B$:
\begin{equation}
  \hat{\bm{z}} = \frac{A-B}{\lVert A-B\rVert},\qquad
  \hat{\bm{x}} = \frac{(N-A) - [(N-A)\cdot\hat{\bm{z}}]\hat{\bm{z}}}{\lVert\cdots\rVert},\qquad
  \hat{\bm{y}} = \hat{\bm{z}}\times\hat{\bm{x}},
\end{equation}
and then
\begin{equation}
  X = B + r\bigl(\cos\theta\,\hat{\bm{z}} + \sin\theta\,(\cos\chi\,\hat{\bm{x}} + \sin\chi\,\hat{\bm{y}})\bigr).
\end{equation}

\subsubsection{Bijective version of the planar local-frame coordinate transformation}

Given $N\ge3$ atom positions where the first three points $\bm{p}_1,\bm{p}_2,\bm{p}_3$ are coplanar reference atoms, \eg atoms in an aromatic ring, a local orthonormal frame is built as
\begin{equation}
  \hat{\bm{e}}_x = \frac{\bm{p}_2-\bm{p}_1}{\lVert\bm{p}_2-\bm{p}_1\rVert},\qquad
  \hat{\bm{e}}_z = \frac{(\bm{p}_2-\bm{p}_1)\times(\bm{p}_3-\bm{p}_1)}{\lVert(\bm{p}_2-\bm{p}_1)\times(\bm{p}_3-\bm{p}_1)\rVert},\qquad
  \hat{\bm{e}}_y = \hat{\bm{e}}_z\times\hat{\bm{e}}_x.
\end{equation}
Each remaining atom $\bm{p}_i$ ($i>3$) is then expressed in this frame as
\begin{equation}
  \bm{q}_i = R^\top(\bm{p}_i - \bm{p}_1),\qquad R = [\hat{\bm{e}}_x,\,\hat{\bm{e}}_y,\,\hat{\bm{e}}_z].
\end{equation}
Since $R$ is an orthogonal matrix that depends only on the reference atoms and the transformation of the remaining atoms is a rigid rotation and translation, the Jacobian determinant of this transformation is unity, \ie, $\log\lvert\det \bm{J}\rvert = 0$.
This transformation is applied to planar groups such as aromatic rings and carboxylate groups.

\subsubsection{The overall bijective transformation and the definition of intermediate representation}

The chignolin protein in this study contains $77$ atoms after removing all hydrogen atoms, giving $231$ Cartesian degrees of freedom.
A preprocessing step fixes the six rigid-body degrees of freedom (three translations and three rotations) by choosing the four reference \ce{C_\alpha} atoms---\ce{C_\alpha^1} (GLY1), \ce{C_\alpha^3} (ASP3), \ce{C_\alpha^8} (THR8), and \ce{C_\alpha^{10}} (GLY10)---with \ce{C_\alpha^8} placed at the origin, \ce{C_\alpha^3} on the positive $y$-axis, \ce{C_\alpha^1} in the $x$-$y$ plane, and \ce{C_\alpha^{10}} on the positive-$z$ side.
After preprocessing, the remaining $225$ internal degrees of freedom are extracted by the bijective transformation described below.

\paragraph*{Basis vectors (6 dimensions).}
Two displacement vectors between reference \ce{C_\alpha} atoms are converted by the bijective spherical coordinate transformation:
the vector from \ce{C_\alpha^8} to \ce{C_\alpha^1}, and the vector from \ce{C_\alpha^1} to \ce{C_\alpha^{10}}.
The transformed vectors are represented as
\begin{equation}
  \ce{C_\alpha^8\bond{-}C_\alpha^1} \mapsto (r_1,a_1,b_1), \qquad
  \ce{C_\alpha^1\bond{-}C_\alpha^{10}} \mapsto (r_2,a_2,b_2).
\end{equation}
Because the preprocessing constrains \ce{C_\alpha^8} to the origin and \ce{C_\alpha^3} to the positive $y$-axis, the component $a_1$ is fixed to zero and is replaced by $y_{\ce{C_\alpha^3}}$, the $y$-coordinate of \ce{C_\alpha^3}.
The six transformed variables are assembled as
\begin{equation}
  \bm{z}_{\text{basis}} = (r_2,\; y_{\ce{C_\alpha^3}},\; r_1,\; b_1,\; a_2,\; b_2),
\end{equation}
with the first two being the CVs -- the distance of \ce{C_\alpha^1\bond{-}C_\alpha^{10}} and \ce{C_\alpha^3\bond{-}C_\alpha^8}.

\paragraph*{Residue frames (60 dimensions).}
The backbone heavy atoms of every residue, the \ce{C_\alpha} and its two bonded neighbors (\ce{N} and \ce{C}), are organized into $10$ groups.
The bijective local-frame coordinate transformation is then applied to all groups, each producing a $9$-variable output: the origin position $(A_x,A_y,A_z)$, three Euler angles $(\alpha,\beta,\gamma)$, and three frame-geometry scalars $(s_1,p,s)$.
The origins of these frames, except \ce{C_\alpha^1}, \ce{C_\alpha^10}, \ce{C_\alpha^3} and \ce{C_\alpha^8} processed above as basis vectors, will be transformed in the next step and are thus dropped here.
As a result, only $6$-variable output per group is retained, yielding $60$ dimensions.

\paragraph*{Torsion coordinates (123 dimensions).}
The bijective torsion transformation is applied to $41$ groups of four sequentially bonded atoms.
First, these comprise $6$ peptide-bond quadruples, $(\ce{N^i},\ce{C_\alpha^i},\ce{C^i},\ce{N^{i+1}})$, that connect consecutive residues along the backbone.
After transformation, each peptide-bond quadruple yields the bond length $r$, the bond angle $\theta$, and the dihedral angle $\chi$, thereby connecting two adjacent residue frames.

Then, $35$ side-chain quadruples extending outward from each residue's \ce{C_\alpha} are transformed via the same bijective torsion transformation.
The $35$ side-chain quadruples follow a hierarchical pattern: the first layer determines atoms bonded directly to the backbone (carbonyl oxygens \ce{O} and first side-chain atoms \ce{C_\beta}), the second layer extends to \ce{C_\gamma}, and deeper layers continue outward to the tips of each side chain.
Each group produces the bond length $r$, bond angle $\theta$, and dihedral angle $\chi$ of the terminal atom, giving a total of $41 \times 3 = 123$ dimensions.

\paragraph*{Planar-group atoms (36 dimensions).}
Aromatic rings and carboxylate groups contain atoms constrained to lie approximately in a plane.
These are handled by $6$ independent applications of the bijective planar local-frame transformation:
\begin{equation*}
  \begin{aligned}
    &\text{TYR2 ring: } (\ce{CG},\ce{CD1},\ce{CE1},\ce{CD2},\ce{CE2},\ce{CZ},\ce{OH});\\
    &\text{TRP9 five-membered ring: } (\ce{CG},\ce{CD2},\ce{CE2},\ce{NE1},\ce{CD1});\\
    &\text{TRP9 six-membered ring: } (\ce{CE2},\ce{CD2},\ce{CE3},\ce{CZ2},\ce{CZ3},\ce{CH2});\\
    &\text{GLY10 terminus: } (\ce{C}_\alpha,\ce{C},\ce{O},\ce{OXT});\\
    &\text{ASP3 carboxylate: } (\ce{C}_\beta,\ce{C}_\gamma,\ce{OD1},\ce{OD2});\\
    &\text{GLU5 carboxylate: } (\ce{C}_\gamma,\ce{C}_\delta,\ce{OE1},\ce{OE2}).
  \end{aligned}
\end{equation*}
As defined previously, in each group, the first three atoms define the reference frame and the remaining atoms are expressed in local Cartesian coordinates.
This produces $12$ local-coordinate vectors of dimension $3$, for a total of $36$ dimensions.

\paragraph*{Reordering and final output.}
We first divide the $60$ dimensions from residue frames into two groups, one is the frame angles $(\alpha,\beta,\gamma)$, the other is the frame geometry $(s_1,p,s)$.
Then, the $41$ torsion dihedral angles $\chi$ are reordered so that the $19$ most physically significant ones---the $6$ peptide-bond dihedrals and $13$ key side-chain dihedrals including \ce{N\bond{-}C_\alpha\bond{-}C_\beta\bond{-}C_\gamma} rotamers---appear first.
The remaining $(r, \theta)$s from the bijective torsion transformation follow these dihedral angles.
Following this, a final overall reordering gives the full $225$-dimensional intermediate representation as
\begin{equation}
  (\bm{s}, \bm{u}) = \bigl(\,
    |\ce{C_\alpha^1\bond{-}C_\alpha^{10}}|,\;
    |\ce{C_\alpha^3\bond{-}C_\alpha^8}|,\;
    \underbrace{\bm{u}_{\text{basis}}}_{4},\;
    \underbrace{\bm{u}_{\text{frame-angle}}}_{30},\;
    \underbrace{\bm{u}_{\text{torsion-}\chi}}_{41},\;
    \underbrace{\bm{u}_{\text{torsion-}r,\theta}}_{82},\;
    \underbrace{\bm{u}_{\text{frame-geometry}}}_{30},\;
    \underbrace{\bm{u}_{\text{planar}}}_{36}
  \,\bigr).
\end{equation}

The overall log-Jacobian determinant of the end-to-end transformation is the sum of the individual contributions:
\begin{equation}
  \log\lvert\det\bm{J}\rvert
  = \bigl(-\log r_1\bigr) + \bigl(-2\log r_2\bigr)
  + \sum_{j=1}^{10}\bigl(-2\log s_{1,j} - \log s_j - \log\lvert\cos\beta_j\rvert\bigr)
  + \sum_{k=1}^{41}\bigl(-2\log r_k - \log\sin\theta_k\bigr),
\end{equation}
where the first two terms are the spherical-coordinate contributions from the two basis vectors,
the second sum corresponds to the local-frame transformations, and the third to the torsion transformations.
Notably, the first basis vector $\bm{r}_1$ is constrained to the $xy$-plane by preprocessing, so its transformation is effectively two-dimensional with Jacobian $-\log r_1$; the second basis vector $\bm{r}_2$ is unconstrained and carries the full three-dimensional spherical Jacobian $-2\log r_2$.
The planar local-frame transformation has a unit Jacobian and does not contribute.

\subsection{Technical details in training}

The classical force field used for the chignolin protein is the \texttt{Amber14} force field combined with the \texttt{GBN2} implicit solvent model.
As above, energies are measured in kJ/mol, so the temperature variable $T$ in VaFES denotes the thermal energy $k_B T_{\mathrm{phys}}$ in kJ/mol. For $350$ K, this gives $T \approx 2.91\si{\kilo\joule\per\mol}$ and therefore $\beta = 1/T \approx 0.35\si{\mol\per\kilo\joule}$.
All hydrogen atoms are removed before applying the bijective transformation, yielding $77$ heavy atoms and $225$ internal degrees of freedom in the intermediate representation.
During energy evaluation, hydrogen atoms are reconstructed from heavy-atom positions by local geometry templates.

The two CVs are inter-\ce{C_\alpha} distances: $|\ce{C}_\alpha^3-\ce{C}_\alpha^8|$ with a range of $[3.4,\,17]$~\AA, and $|\ce{C}_\alpha^1-\ce{C}_\alpha^{10}|$ with a range of $[4.5,\,10]$~\AA.
During training, the two CVs are uniformly sampled over their respective ranges.
The prior distribution for the $223$-dimensional auxiliary variables is a truncated Gaussian whose mean and deviation are learnable parameters.

The chignolin application employed the same cubic-spline flow as in the previous cases, except that we used a more capable parameter neural network;
the spline parameters in each coupling layer are produced by a joint MLP with \texttt{ResNet1d}~\cite{resnet}.
This network first maps the input through a MLP to a high-dimensional vector, which is reshaped into a multi-channel 1D signal.
The signal passes through an initial 1D convolution, followed by a stack of residual 1D-convolutional blocks with bottleneck structure, and finally through point-wise fully connected layers to produce the spline parameters.
Each residual block consists of a $1\times 1$ projection to a narrower hidden channel width, a 1D convolution, and a $1\times 1$ expansion back to the full width.

Using different partition schemes, the coupling structure proceeds in three stages.
In the first stage, there are three sequential coupling layers with increasing size: dimensions $2$--$5$ (basis variables), then $6$--$41$ (frame angles and significant torsions), then $42$--$54$ (remaining key torsions), each conditioned on all previously transformed dimensions.
In the second stage, $14$ coupling layers with a typical checkerboard partition iterating between them,
mixing all degrees of freedom.
A single coupling layer then transforms dimensions $55$--$224$ conditioned on all preceding dimensions of $2$--$54$.
In the final stage, $8$ coupling layers with another typical checkerboard partition over all $223$ auxiliary dimensions provide global mixing.

During training, to overcome local minima and metastable states, we used the VaTD~\cite{vatd} training scheme with the inverse temperature $\beta$ as the external parameter, sampled continuously over $[0.1, 0.6]$.
The remaining hyperparameters are listed in Table~\ref{tab:chig}.

\begin{table}[h]
\centering
\caption{VaFES hyperparameters for the chignolin application.}
\label{tab:chig}
\begin{tabular}{l l}
\hline\hline
Parameter & Value \\
\hline
Temperature $\beta$ & $0.35\si{\mol\per\kilo\joule}\ (\approx 350\,\mathrm{K})$ \\
Spline bins $K$ & $32$ \\
Coupling layers (stage 1 / 2 / 3) & $3$ / $14 + 1$ / $8$ \\
MLP hidden dims & $(512,\, 1024,\, 1024)$ \\
\texttt{ResNet1d} channels / hidden channels & $256$ / $128$ \\
\texttt{ResNet1d} kernel size & $25$ \\
\texttt{ResNet1d} residual conv layers & $8$ \\
\texttt{ResNet1d} point-wise FC layers & $4$ \\
\texttt{ResNet1d} hidden FC width & $128$ \\
Activation & ELU \\
Prior distribution & Truncated Gaussian \\
Optimizer & Adamax \\
Learning rate & $2\times 10^{-4}$ \\
LR schedule & StepLR (step $1000$, $\gamma=0.8$) \\
Batch size & $128$ \\
\hline\hline
\end{tabular}
\end{table}

\end{document}